\documentclass[aps,amsmath,amssymb,prb,twocolumn,superscriptaddress,showpacs,floatfix, longbibliography, reprint]{revtex4-2}
\usepackage{amsmath}
\usepackage{bbold}
\usepackage{braket}
\usepackage{graphicx}
\usepackage{booktabs}
\usepackage{array}
\usepackage[T1]{fontenc}
\usepackage{bm}
\usepackage{color}
\usepackage{xcolor}
\usepackage{makecell}
\usepackage[colorlinks=true, urlcolor=blue, linkcolor=red, citecolor=blue]{hyperref}
\usepackage[caption=false]{subfig}
\begin{document}

\title{Variational study of the magnetization plateaus in the spin-1/2 kagome Heisenberg antiferromagnet: An approach from vision transformer neural quantum states}

\author{Andreas Raikos}
\email{andreas.raikos@irsamc.ups-tlse.fr}
\affiliation{Universit\'e de Toulouse, CNRS, Laboratoire de Physique Th\'eorique, Toulouse,  France.}
\author{Sylvain Capponi} 
\email{sylvain.capponi@irsamc.ups-tlse.fr}
\affiliation{Universit\'e de Toulouse, CNRS, Laboratoire de Physique Th\'eorique, Toulouse,  France.}
\author{Fabien Alet}
\email{fabien.alet@cnrs.fr}
\affiliation{Universit\'e de Toulouse, CNRS, Laboratoire de Physique Th\'eorique, Toulouse,  France.}

\date{\today}
\begin{abstract}
    We analyze the magnetization curve of the spin-1/2 kagome Heisenberg model in a magnetic field. Using state-of-the-art variational wave functions based on neural networks, we confirm the presence of robust magnetization plateaus at $m=1/3$, $5/9$, and $7/9$ of the saturation value, stabilized by a spontaneous symmetry breaking of lattice translations with a $\sqrt{3}\times \sqrt{3}$ unit cell. Regarding the more challenging $m=1/9$ plateau, we find two competing valence bond crystals depending on the system size, both breaking translation as well as point group symmetries and with a larger $3\times 3$ unit cell. Such quantum states with local modulations of the magnetization average values could be observed experimentally in the near future.
\end{abstract}

\maketitle

\section{Introduction}\label{sec:introduction}

Amongst strongly correlated systems, frustrated quantum magnets offer a unique and dual playground as they exhibit rich new physics including unconventional phases of matter at low energy, while posing several challenges to theoretical, numerical and experimental approaches to describe them. Unlike their non-frustrated counterparts, several frustrated quantum spin systems do not display long-range magnetic order at low temperature. The nature of the ground state of frustrated quantum spin systems displays a rich variety of nonmagnetic states, including quantum spin liquids (QSL)~\cite{Savary2017,KnolleMoessner,Capponi2025}---with or without a gap above the ground state, valence bond crystals~\cite{MisguichLhuillier}, spin nematics~\cite{Starykh_2015}, spin ices~\cite{Bramwell_2020}, etc.

One of the most celebrated yet unsolved cases in frustrated quantum magnetism is the spin-1/2 Heisenberg antiferromagnet on the two-dimensional (2d) kagome lattice made of corner sharing triangles. The nature of its ground state is not fully resolved despite years of intensive studies, both on the theoretical and experimental fronts (for reviews, see Refs.~\onlinecite{LacroixMendelsMila_book,zhu_quantum_2025,Mendels_2016}).
A slightly simpler, albeit equally interesting, situation occurs when the kagome Heisenberg $S=1/2$ antiferromagnet is submitted to an external magnetic field $h$. There, the behavior of the magnetization per spin $m$ is not smooth with the field, and magnetization plateaus occur for particular rational values of $m$. This is a rather generic situation in frustrated quantum antiferromagnets~\cite{Honecker2004}, which can be easily understood with a simple bosonic picture~\cite{Matsubara1956} where spins up (respectively down) behave as hardcore bosonic particles (respectively holes). On frustrated lattices, particles generically obtain a reduced kinetic energy, hence, for commensurate fillings, i.e., fractional $m$, a superfluid/insulator transition can occur due to a dominant repulsive interaction energy over the kinetic one~\cite{MomoiTotsuka2000}.

Magnetization plateaus have been extensively studied on the kagome lattice. First, an exact magnon crystal has been found at $m=7/9$, above which there is a metamagnetic jump to saturation~\cite{Schulenburg2002}. Physically, it corresponds to a valence bond crystal (VBC), a nonmagnetic state that spontaneously breaks translation symmetry, where some spins are fully polarized while the remaining hexagons correspond to the quantum superposition of a single down spin delocalized on each hexagon, see Fig.~\ref{fig:triple:7_9_observables}. Later studies have revealed the existence of additional robust plateaus for $m=5/9$ and $1/3$, using exact diagonalization (ED)~\cite{Capponi2013}, density-matrix renormalization-group (DMRG)~\cite{Nishimoto2013}, or infinite projected entangled pair states (iPEPS)~\cite{Picot2016,Chen2018,Okuma2019} etc. Most studies indicate similar VBCs with a $\sqrt{3}\times \sqrt{3}$ pattern, i.e., nine-site unit cell, with almost fully polarized spins and resonating hexagons having respectively 2 or 3 down spins per hexagon. Note that a recent variational Monte Carlo (VMC) study finds a different type of VBC~\cite{He_2025}. 

The possibility of an additional plateau at $m=1/9$ has emerged more recently~\cite{Nishimoto2013,Picot2016,Chen2018}, although its precise nature remains under debate. DMRG calculations and a recent VMC study based on fermionic partons point towards a topological $\mathbb{Z}_3$ quantum spin liquid~\cite{Nishimoto2013, He2024_one_nine_Z3_QSL}, while iPEPS studies indicate  a more conventional 18-fold degenerate VBC~\cite{Picot2016} or a VBC with gapless excitations and a $\sqrt{ 3 } \times \sqrt{ 3 }$ unit cell~\cite{Fang2023}. Complementary ED and limiting perturbative arguments starting from a distorted kagome lattice have also  advocated for a $\sqrt{ 3 } \times \sqrt{ 3 }$  VBC~\cite{Morita2024}. Finally, a recent VMC study based on a general resonating valence bond ansatz has found evidence of a VBC with a larger,  $3\times 3$ unit cell characterized by a windmill-shaped motif~\cite{Cheng2026}. 

On the experimental side, this topic has also been very active recently thanks to advances in the generation of high magnetic fields as well as newly synthesized quantum magnetic compounds, see Ref.~\onlinecite{Yoshida2022} for a review. Until recently, only the $m=1/3$ plateau had been observed~\cite{Okamoto2011,Hiroi2015,Yoshida2017,Kato_one-third_2024} among all the expected ones. The $m=1/9$ plateau has been finally observed in 2024~\cite{Jeon2024,Suetsugu2024} in yttrium-based kagome materials, following their discovery.

To capture ground states of complex quantum many-body systems, a powerful recent theoretical improvement involves the use of neural quantum states (NQS), that is, variational wave functions parametrized as neural network architectures. Since their introduction in 2017 \cite{Carleo2017}, NQS have been used to study a wide variety of physical systems, ranging from frustrated quantum magnets, for which they constitute a bona fide state-of-the-art variational method,  to models with fermionic 
and bosonic
degrees of freedom \cite{Lange2024, Y_Pei_2024, Denis2025}.
Their remarkable success can be attributed both to the excellent expressivity of neural networks, encapsulated in universal approximation theorems \cite{Cybenko1989,Hornik1989}  which guarantee that sufficiently  parametrized neural networks can in principle represent arbitrary (wave)functions, along with progress in optimizers \cite{Sorella1998, Amari1998, Becca_2017, Martens2020, Chen2024,Rende2024} that have rendered practical the training of NQS with millions of parameters.

In this work, we employ a very recently proposed architecture for NQS, known as Vision Transformers (ViT) with a factored attention mechanism~\cite{Viteritti2025,Rende2024_potts,Rende2025_QK}. Originally introduced in the machine learning community for computer vision tasks~\cite{Dosovitskiy2021}, the ViT architecture has been recently adapted as a variational state~\cite{Viteritti2023} and has been shown to achieve state-of-the-art performance as an ansatz for frustrated quantum magnets, such as the  $J_1$ - $J_2$ Heisenberg  or Shastry-Sutherland models~\cite{Viteritti2023,Rende2024, Viteritti2025}. More recently, ViT-based NQS have enabled the transfer of additional state-of-the-art machine learning techniques to the VMC framework, such as fine-tuning~\cite{Rende2024_ft} and the foundation-model paradigm~\cite{Rende2025}. The so-called factored attention variant allows one to enforce translation invariance in NQS wave functions, which recent studies suggest is key for the strong performance of the  model~\cite{Rende2025_QK,Nutakki2025}.

Using this architecture, we study in this work the magnetization plateaus of the kagome Heisenberg antiferromagnet, with a particular focus on determining the nature of the challenging $m=1/9$ plateau state. The plan of the manuscript is as follows. In Sec.~\ref{sec:model}, we first introduce the kagome Heisenberg model, the NQS architecture chosen to study its magnetization plateaus, as well as useful symmetry considerations for the lattices considered in this work. We then start the presentation of our NQS results in Sec.~\ref{sec:results} by the magnetization curve for two lattice sizes, which clearly displays plateaus at the expected values. We then consider each magnetization plateau, by order of increased complexity. We analyze the $m=7/9$ plateau in Sec.~\ref{sec:7}, $m=5/9$ in Sec.~\ref{sec:5}, and $m=1/3$ in Sec.~\ref{sec:3}, where we obtain results consistent with most previous studies. This allows us to confirm the validity of our approach (we systematically find variational energies lower than or equal to the best known values), including the symmetry analysis. We finally consider the $m=1/9$ plateau in Sec.~\ref{sec:9}, where our variational results point towards the existence of two valence bond crystals with slightly different symmetry contents which are competing at low energy. Finally, Sec.~\ref{sec:conc} provides a critical analysis and comparison of our $m=1/9$ results to those obtained with other approaches, as well as perspectives on the use of the ViT NQS for other frustrated quantum spin systems.

\section{Model and Method}\label{sec:model}
\subsection{Heisenberg model}
\begin{figure}[t]
    \centering
    
    \begin{minipage}[t]{0.48\columnwidth}
    \centering
    \includegraphics[width=1.1\linewidth,
    trim = 50 25 60 36,
    clip]{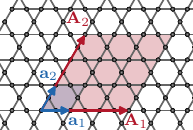}
    \end{minipage}
    \hfill
    \begin{minipage}[t]{0.48\columnwidth}
        \centering
        \includegraphics[width=\linewidth]{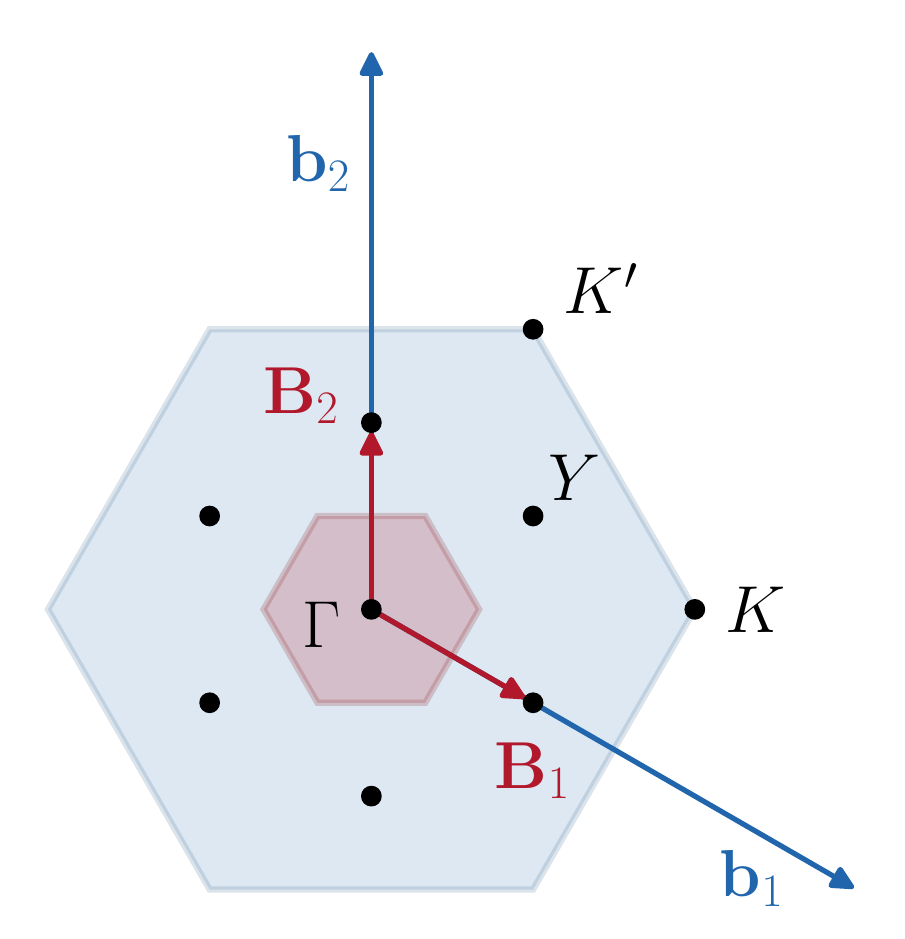}
    \end{minipage}
    \caption{Kagome lattice and ViT patch geometry. Blue (red) arrows indicate the lattice (patch-superlattice) primitive vectors and the corresponding primitive unit cell (superlattice unit cell, i.e., the $3\times 3$ ViT patch). The choice of patch induces folding of the original Brillouin zone (blue) into the reduced superlattice zone (red). The nine marked points are the momenta in the $\mathcal{K}_\Gamma$ set [see Eq.~(\ref{eq:vit_momenta})] representable by the ViT for this patch geometry.}
    \label{fig:lattice_and_bz}
\end{figure}

We study the antiferromagnetic spin-1/2 Heisenberg model on the kagome lattice (Fig.~\ref{fig:lattice_and_bz}) with a magnetic field
\begin{equation}\label{eq:model}
 {\cal H} = J\sum_{\langle ij\rangle} \boldsymbol{S}_i\cdot \boldsymbol{S}_j -h\sum_i S_i^z,
\end{equation}
where the first sum runs over nearest-neighbor bonds, $J=1$ is taken as the unit of energy and $h$ is the magnetic field along the $z$ direction. 
We define the magnetization per site $m=2 \langle S^z \rangle/N$,
where $S^z$ is the conserved $z$-component of the total spin ($S^z=\sum_i S_i^z$) and $N$ the number of sites, which ensures a saturation value $m=1$. The unit cell of the kagome hosts $n_c=3$ sites, such that $N=3L^2$. We will primarily deal with samples with $L=6,9$ ($N=108, 243$) with periodic boundary conditions. 

In the absence of the magnetic field, the nature of the ground state is still highly debated, and is often thought to be a quantum spin liquid (with short-range spin-spin correlations) but its precise nature (gapped vs gapless, topological etc) is not settled~\cite{zhu_quantum_2025}. In a finite magnetic field, the situation is somehow simpler for some magnetizations $m$ with the appearance of incompressible phases leading to finite magnetization plateaus in the magnetization curve, as discussed in Sec.~\ref{sec:introduction}.
Quite interestingly, a featureless unique gapped ground state is possible on a given plateau $m$ iff $n_c\,S(1-m) \in \mathbb{N}$ where $n_c$ is the number of sites per unit cell, $S$ the spin value~\cite{OYA1997,Hastings2004}. In our case, we have $n_c=3$ and $S=1/2$ and the existence of magnetization plateaus at certain fractional values such as $1/9$, $5/9$ or $7/9$, necessarily implies ground-state  degeneracy or gapless excitations. Conversely, for $m=1/3$, a featureless unique gapped ground state is possible, an illustration of which has been constructed in a bosonic model~\cite{Parameswaran2013}. However, this possibility does not occur for the kagome model at $m=1/3$, where a VBC that spontaneously breaks lattice symmetry has been found instead~\cite{Cabra2005,Nishimoto2013,Capponi2013,Picot2016}.

\subsection{Vision transformer NQS}\label{sec:vit}

For a quantum system of $N$ spin-$1/2$ sites, NQS implement a map from spin configurations $\boldsymbol{\sigma}\in\left\{-1,1\right\}^{N}$, representing computational basis states $\ket{\boldsymbol{\sigma}}=\ket{\sigma_{1}^{z},\ldots,\sigma_{N}^{z}}$ with $\sigma_{i}^{z}=\pm1$, to the amplitudes $\Psi_{\boldsymbol{\theta}}\left( \boldsymbol{\sigma} \right)=\langle \boldsymbol{\sigma} | \Psi_{\boldsymbol{\theta}} \rangle$ of the $\boldsymbol{\theta}$-parametrized  variational wave function $\ket{\Psi_{\boldsymbol{\theta}}}=\sum_{\boldsymbol{\sigma}}\Psi_{\boldsymbol{\theta}}\left( \boldsymbol{\sigma} \right)\ket{\boldsymbol{\sigma}}$.

\begin{figure}
  \centering
  \includegraphics[width=1\linewidth]{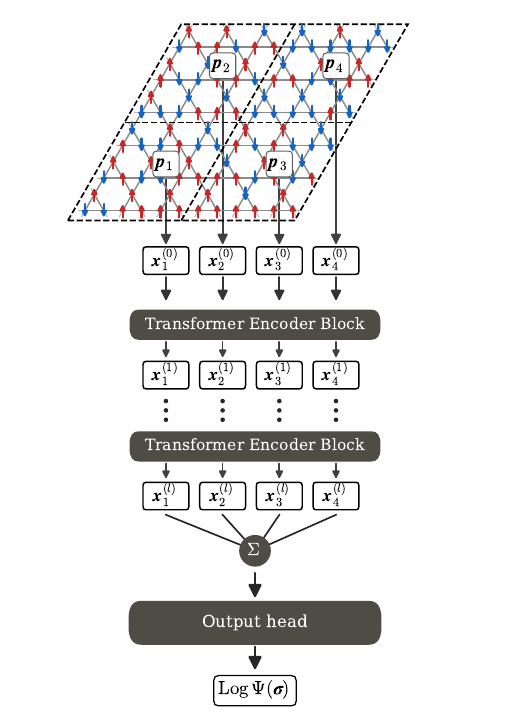}
  \caption{Schematic representation of the Vision transformer NQS described in Sec. \ref{sec:vit}. A spin configuration on the kagome lattice is divided into patches $\boldsymbol{p}_i$, which are mapped to embedding vectors $\boldsymbol{x}_i^{(0)}$. The embeddings are then processed by $l$ translationally equivariant transformer encoder blocks, resulting in output vectors $\boldsymbol{x}^{(l)}$. These are sum-pooled and processed by a shallow output layer to produce the log-amplitude. See Ref.~\cite{Viteritti2025} for more details on the architecture.}
  \label{fig:vit_schematic}
\end{figure}

In this work, we use a two-component Vision Transformer (ViT) NQS architecture as introduced in Ref. \cite{Viteritti2025}, which employs a real-valued deep ViT encoder followed by a shallow fully-connected output layer with complex weights. We refer to Refs.~\cite{Viteritti2025,Rende2025_QK, Viteritti2023} for motivations and implementation details for this architecture, and just briefly mention here the ingredients that are useful for later analysis. This architecture implements the $\ket{\boldsymbol{\sigma}}\rightarrow \Psi_{\boldsymbol{\theta}}\left( \boldsymbol{\sigma} \right)$ mapping in a three-stage process: 
(i) The spin configuration $\boldsymbol{\sigma}$ is first partitioned into a sequence $\left( \boldsymbol{p}_1, \ldots, \boldsymbol{p} _n\right)$ of  $n=N/P$ patches, each containing $P$ spins. The patch size and geometry are lattice- and problem-dependent, and we will discuss our specific choice next.  Each patch $\boldsymbol{p}_i$ is then linearly mapped to an embedding vector $\boldsymbol{x}_{i} \in \mathbb{R}^{d}$, resulting in a sequence $( \boldsymbol{x}_{1}^{(0)},\ldots,\boldsymbol{x}_{n}^{(0)} )$. The integer $d$ is the embedding dimension and is a hyperparameter of the model;
(ii) The sequence of embeddings is then passed through $l$ transformer encoder blocks, each  consisting of a  translationally equivariant factored attention layer \cite{Rende2024_potts} with $n_h$ attention heads followed by a two-layer fully connected network, combined with prenorm skip connections. In this factored attention variant, the query-key dot product of the standard attention mechanism \cite{Vaswani} is replaced by learnable attention weights that depend only on the relative displacement between patches. As a result, the combined embedding-encoding stack is rendered equivariant under patch-level  translations of the input spin configuration; 
(iii) Finally, the {output sequence $( \boldsymbol{x}_{1}^{(l)},\ldots, \boldsymbol{x}_{n}^{(l)} )$  of the last encoder block} is sum-pooled over the patch index into a single vector {$\boldsymbol{z}=\sum_{i}\boldsymbol{x}_{i}^{(l)}$} and mapped through a  single fully-connected layer \cite{Carleo2017} to the (natural) log-amplitude:
    \begin{equation}
    \log \Psi\left( \boldsymbol{\sigma} \right)  =\sum_{i=1}^{d}\log \cosh \left( b_{i} + \sum_{j=1}^d W_{ij}z_{j} \right) 
        \end{equation}
with trainable parameters $b \in \mathbb{C}^{d}, W\in \mathbb{C}^{d\times d}$. See Fig. \ref{fig:vit_schematic} for a schematic illustration of the architecture. 

Through the sum-pooling operation, the final layer
augments the patch-translation equivariance property of the embedding-encoder stack into full patch-translation invariance for the total ViT NQS. This means that the model enforces the property  $\Psi \left( T_{\boldsymbol{R}_{p} }\boldsymbol{\sigma} \right)= \Psi \left( \boldsymbol{\sigma} \right)$, with $T_{\boldsymbol{R}_p}$ being the patch-translation operator. Such wave functions  correspond to the $\Gamma$ point of the  folded Brillouin zone associated with the superlattice defined by the patch tiling. Wave functions with nonzero superlattice momentum can also be obtained with a minor modification of the output layer \cite{Viteritti2025}.

The ViT NQS can therefore accommodate both phases that respect the translation symmetry of the physical lattice, such as QSL states, and also ones that break translation symmetry, provided that the patch is chosen to be commensurate with the enlarged unit cell. All results that follow are obtained using a patch of $P=27$ spins arranged as a $3\times3$ supercell of kagome primitive unit cells, see Fig. {\ref{fig:lattice_and_bz}}. This choice is compatible with both the putative QSL as well as all VBC states previously reported for the magnetization plateaus of the kagome Heisenberg antiferromagnet. 

The ansatz $\ket{\Psi_{\boldsymbol{\theta}}}$ contains variational parameters
$\boldsymbol{\theta}$ that are optimized by minimizing the variational energy $\displaystyle 
E_{\boldsymbol{\theta}}
=\frac{\langle \Psi_{\boldsymbol{\theta}}|\mathcal{H}|\Psi_{\boldsymbol{\theta}}\rangle}
{\langle \Psi_{\boldsymbol{\theta}}|\Psi_{\boldsymbol{\theta}}\rangle}$ through the standard VMC framework \cite{Becca_2017}, using Markov chain Monte Carlo sampling. The energy minimization is performed iteratively using the Subsampled Projected-Increment Natural Gradient Descent (SPRING) algorithm \cite{Goldshlager2024}
 which builds upon the Stochastic Reconfiguration (SR) \cite{Sorella1998, Amari1998,Martens2020}, and MinSR  \cite{Chen2024, Rende2024} schemes. 
 
\subsection{Symmetries}
\label{sec:sym}

In order to understand the potential symmetry breakings occurring in a ground state at finite magnetization, it is useful to consider its transformations under various lattice symmetries. As described above, we obtain wave functions which are invariant under patch translation, but we can further obtain information on their transformations under point-group symmetries.

The choice of patch for the ViT embedding layer defines a superlattice of the original kagome lattice, with superlattice vectors $\mathbf A_{1}=m_{1}\mathbf a_{1}$ and $\mathbf A_{2}=m_{2}\mathbf a_{2}$, where $\mathbf a_{1},\mathbf a_{2}$ are the kagome primitive lattice vectors and  $m_1, m_2$ are the number of primitive unit cells in each direction of the patch. Denoting by $\mathbf{b}_{1}, \mathbf{b}_{2}$ ($\mathbf B_{1}, \mathbf B_{2}$) the reciprocal lattice vectors of the kagome lattice (superlattice), a momentum  vector $\mathbf{k}$ in the original Brillouin zone (BZ)  folds onto a superlattice Brillouin zone (SBZ) momentum $\boldsymbol{\kappa}$ via
$
\mathbf{k} =\boldsymbol{\kappa} + n_{1} \mathbf{B}_{1} + n_{2} \mathbf{B}_{2},
$
with $n_1, n_2 \in \mathbb{Z}$ (see Fig.~\ref{fig:lattice_and_bz}).

Since the states represented by the  ViT architecture are invariant under patch translations, they correspond  to the $\boldsymbol{\kappa} =0$ (i.e., $\Gamma$) point of the SBZ. For our choice of  patch with $m_1=m_2=3$, the set of BZ momenta mapped to this point through folding is given as

\begin{equation}\label{eq:vit_momenta}
    \mathcal{K}_{\Gamma}= \left\{\frac{p}{3}\mathbf{b}_{1}+\frac{q}{3}\mathbf{b}_{2}\,|\,p,q\in\{0,1,2\}\right\}. 
\end{equation}
This includes the three high-symmetry $\Gamma, K, K^\prime$ points of the BZ, along with six internal, mirror-symmetric BZ points $\{Y_i\}_{i=0}^5$ related by $\pi/3$ rotations (see Fig. \ref{fig:lattice_and_bz}). 

The ViT with our chosen patch can thus represent  superpositions

\begin{equation}
    \ket{\Psi_{\mathrm{ViT}}}= \sum_{\mathbf{k}\in\mathcal{K}_{\Gamma}} c_{\mathbf{k}}\ket{\psi_{\mathbf{k}}}
\end{equation}
where $\ket{\psi_{\mathbf{k}}}$ denotes a momentum eigenstate of the original kagome lattice and $c_{\mathbf{k}}\in \mathbb{C}$. We note that for our choice of samples with $L=6,9$, all nine folded momenta are present in the finite-size BZ.

Since the Hamiltonian (\ref{eq:model}) commutes with all elements of the (finite-size) lattice space group $G$, its eigenstates can be labeled by irreducible representations (irreps) $\alpha$ of $G$. Each space group irrep can be specified by an equivalence class of BZ momenta, the so-called star $[\mathbf{k}]$ (henceforth denoted  by a representative momentum $\mathbf{k}$) together with  an irrep of its associated little cogroup $G_{\mathbf{k}}$, a subgroup of the kagome lattice point group $C_{6v}$ (see Appendix \ref{sec:App-symmetry} for definitions and conventions). For symmetry-breaking states such as the VBCs expected at magnetization plateaus, finite clusters nevertheless exhibit quasidegenerate low-energy states belonging to different irreps, which become degenerate in the thermodynamic limit, see, e.g., Ref.~\onlinecite{MisguichLhuillier}. In practice, we find that on the magnetization plateaus, the ViT NQS converges
 to symmetry-broken representative states with support within this quasidegenerate manifold. 
 
 To characterize the irrep content of such states, we utilize two complementary symmetry decomposition approaches that we describe now. From a numerical standpoint, once we obtain a variational state $\ket{\Psi_m}$ with low energy, we can first construct the projector onto a given space-group irrep $\alpha$ as
\begin{equation}\label{eq:irrep_projector}
P_{\alpha}=\frac{d_\alpha}{|G|}\sum_{g\in G}\chi_{\alpha}(g)^{*}\,U(g),
\end{equation}
where $d_\alpha$ and $\chi_{\alpha}(g)$ are the dimension and character of $\alpha$, and $U(g)$ denotes the representation of the group element $g$. We then estimate, using Monte Carlo sampling, the weight

\begin{equation}\label{eq:irrep_weight}
w_{\alpha}
=\frac{\langle \Psi_m| P_{\alpha}|\Psi_m\rangle}{\langle \Psi_m|\Psi_m\rangle},
\end{equation}
which quantifies the support of the state $\ket{\Psi_m}$ in the corresponding irrep. By construction, the weights satisfy $\sum_{\alpha} w_\alpha \simeq 1$ within statistical error. In order to compare with different symmetry-broken candidate states (such as different VBCs) described in the second approach below, it is convenient to define the rescaled weights 

\begin{equation}\label{eq:rescaled_weights}
\tilde{N}_{\alpha} \equiv D w_{\alpha},
\end{equation}
with $D$ being the degeneracy of the corresponding candidate ground-state manifold. 

To obtain the predicted degeneracies for VBCs, we can obtain a purely group-theoretic prediction for the irrep content by starting from an idealized, symmetry-broken VBC state $\ket{\Phi_{m}}$ for each plateau, inferred from the local observables of the corresponding optimized state $\ket{\Psi_{m}}$ (e.g., as shown in Fig.~\ref{fig:observables_high_field_plateaus}). To this end, we determine the stabilizer $S_m \subset G$, that is, the subgroup of space-group elements $s \in S_m$ whose representations satisfy $U\left( s \right)\ket{\Phi_{m}}=\ket{\Phi_{m}}$. Assuming that the action of the full space group $G$ on $\ket{\Phi_{m}}$ generates $D=\left|G\right|/\left|S_m\right|$ separate, linearly independent states spanning the degenerate ground-state manifold of the VBC, the multiplicity $n_{\alpha}$ of each irrep can be obtained from the character-stabilizer formula \cite{Wietek2017}:
\begin{equation}\label{eq:character_stabilizer}
n_{\alpha}=\frac{1}{\left|S_m\right|} \sum_{s\in S_m} \chi_{\alpha} \left( s \right)^* .
\end{equation}
For direct comparison with the numerically-obtained (rescaled) weights (\ref{eq:rescaled_weights}) of the first approach, we denote as $N_\alpha \equiv d_\alpha n_\alpha$  the dimension of the subspace of the  VBC ground-state manifold that transforms according to the $d_{\alpha}$-dimensional irrep $\alpha$.

We note that Eq.~(\ref{eq:character_stabilizer}) provides a simplified counting rule which, while useful, may ignore additional internal structure of the low-energy quasidegenerate subspace on a finite cluster. In particular, this subspace may not be generated by a single idealized VBC state with trivial stabilizer action, which may lead to size-dependent changes in the predicted irrep decomposition. Such a situation arises, for example, in the case of the exact magnon crystal state at the $m=7/9$ plateau, where, for odd $L$,  the wave function acquires a minus sign under certain stabilizer elements due to the internal symmetry of the VBC. This results in a size-dependent interchange of irreps, a simple illustration of which is provided for the $m=7/9$ plateau in Sec.~\ref{sec:7}. 

\subsection{Implementation details} \label{sec:implementation}

We use ViT NQS with $d=160$, $n_h=40$ and $l=4$, corresponding to  approximately $N_p \simeq 1.1$ million trainable parameters. For the optimization runs, we used (unless otherwise stated, see Appendix~\ref{sec:App-protocol}) a cosine decaying learning rate~\cite{Loschilov} $\tau$ from $0.03$ to $10^{-3}$, cosine decaying diagonal shift regularization $\lambda$ from $0.01$ to $10^{-4}$ and SPRING momentum~\cite{Goldshlager2024} $\mu =0.9$. Monte Carlo sampling (with a typical number of Monte Carlo samples $M=8192$) is performed within fixed $S^z$ magnetization, using update proposals that exchange pairs of opposite spin orientations. We also tested sampling with updates that change the total magnetization, but, for the chosen architecture, we observed no convergence benefit and found in general the optimization to be less stable.  All simulations were performed using the NetKet library~\cite{netket3_2022, netket2_2019}.

\section{Results}\label{sec:results}

\begin{figure}[t]
\includegraphics[width=\columnwidth]{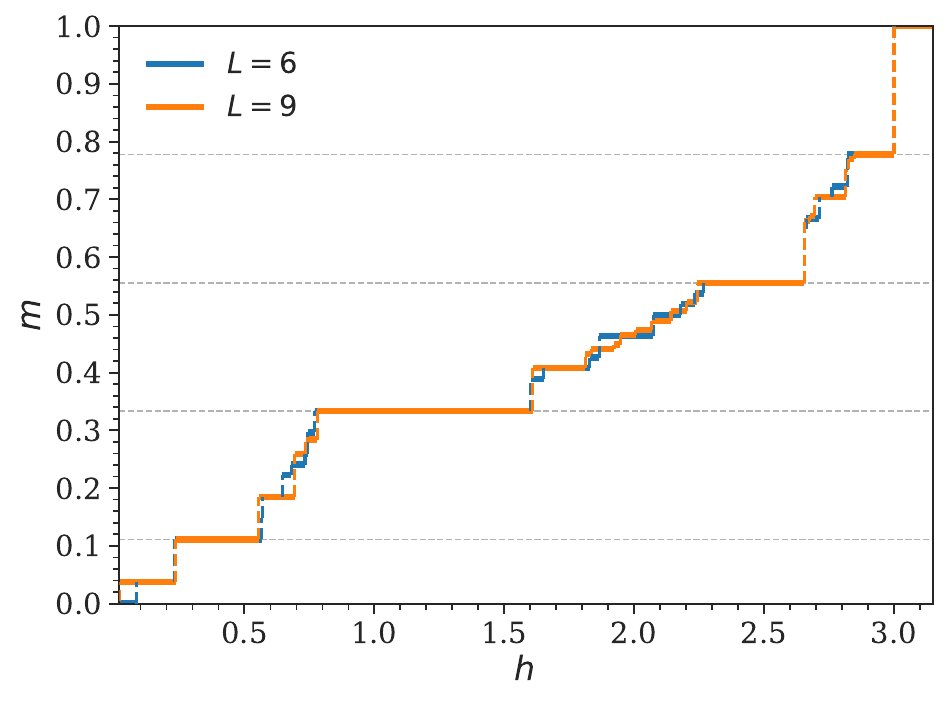}
\caption{Magnetization curve for the spin-1/2 kagome Heisenberg antiferromagnet with $L=6$ and $9$ lattice unit cells per linear dimension. The dotted lines correspond to the $m=1/9$, $1/3$, $5/9$, and $7/9$ plateaus.}
\label{fig:magnetization_curve}
\end{figure}

\begin{table}
  \caption{\label{tab:plateau_widths} Field ranges of the magnetization plateaus for the $L=6$ and $L=9$ kagome clusters.} 
  \begin{ruledtabular}
    \begin{tabular}{lcccc}
      $m$ & 1/9 & 1/3 & 5/9 & 7/9 \\ \colrule
      $L=6$ & [0.23, 0.57] & [0.78, 1.60] & [2.27, 2.65]  & [2.82, 3]  \\
      $L=9$ & [0.24, 0.56] & [0.78, 1.61]  & [2.25, 2.65] & [2.84, 3]    \\
    \end{tabular}
  \end{ruledtabular}
\end{table}

\begin{figure*}[t] 
  \centering
  \subfloat[\large \vspace{-6mm} $\boldsymbol{m=1/3}$]{
    \includegraphics[
      width=0.3\textwidth,
      trim=36mm 40mm 36mm 45mm, 
      clip
    ]{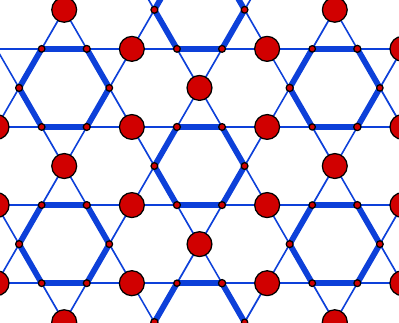}
    \label{fig:triple:1_3_observables}
  }
  \hfill
  \subfloat[\large \vspace{-6mm} $\boldsymbol{m=5/9}$]{
    \includegraphics[
      width=0.3\textwidth,
      trim=36mm 40mm 36mm 45mm,
      clip
    ]{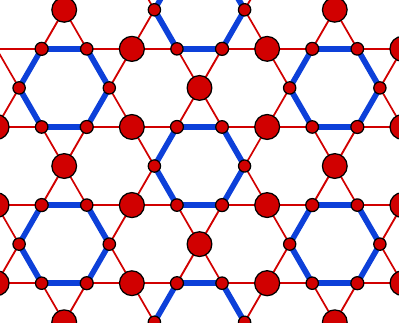}
    \label{fig:triple:5_9_observables}
  }
  \hfill
  \subfloat[\large \vspace{-6mm} $\boldsymbol{m=7/9}$]{
    \includegraphics[
      width=0.3\textwidth,
      trim=36mm 40mm 36mm 45mm,
      clip
    ]{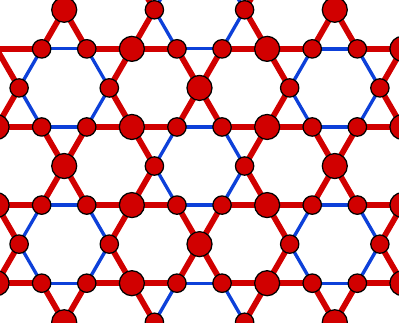}
    \label{fig:triple:7_9_observables}
  }
  \vspace{5mm}
  \caption{Magnetization per site and bond energy   (see Eq.~\eqref{eq:mag_bond_energy})  for the VBC states at the three high-field plateaus $m= 1/3$, $5/9$, and $7/9$. Disk diameter and bond width encode the magnitude of the observables as estimated directly from $2^{16}$ Monte Carlo samples, while red (blue) indicates a positive (negative) sign. Spatially averaged expectation values and standard deviations are reported in Tables \ref{tab:magnetization}  and \ref{tab:bond_energies}. }
  \label{fig:observables_high_field_plateaus}
\end{figure*}

We consider kagome lattice samples with the geometry illustrated in Fig. \ref{fig:lattice_and_bz}, with periodic boundary conditions along both directions. Assuming $L$ lattice unit cells per linear dimension, the resulting clusters are rhombic tori 
with a total of $N=3L^{2}$ spins per sample. We focus on the cases  $L=6 \left( N=108 \right)$ and $L=9\left( N=243 \right)$, both of which are compatible with  $\sqrt{ 3 }\times \sqrt{ 3 }$ order and can therefore accommodate the VBC states expected to be associated with the $m=1/3, 5/9$ and $7/9$ magnetization plateaus. 

\subsection{{Magnetization curve}}

To construct the zero-temperature magnetization curve $m(h)$, we optimize our NQS within   all fixed-magnetization sectors {for the $L=6$ and $L=9$  samples} and obtain  the within-sector ground-state energies $E_0\left( m \right)$ at zero field. The corresponding field-dependent energy is given by $E_{h}\left( m \right)=E_0\left( m \right) - \frac{N}{2}hm$, while the magnetization of the system at field $h$ is obtained as the minimization $m\left( h \right) =\arg\min_{\tilde{m}}  \left\{E_0\left( \tilde{m} \right) - \frac{N}{2}h\tilde{m}\right\}.$
In practice, we calculate $m(h)$ by finding the lower convex hull of the set of points $(m,E_0)$ for all magnetization sectors. 

The obtained magnetization curves are displayed in Fig.~\ref{fig:magnetization_curve} and are consistent with the existence of four plateaus at magnetizations $m=1/9, 1/3, 5/9$ and $7/9$. The obtained widths of the plateaus are given in Table \ref{tab:plateau_widths}. Let us emphasize that we have not put as much effort in the numerical simulations \emph{outside} these plateaus (typically, we have performed only half the number of iterations) so that our results on the full magnetization curve are more qualitative than on these specific plateaus. In particular, there are other smaller size finite-size plateaus that presumably will not persist in the thermodynamic limit.  
That being said, we can already identify several robust features: (i) the widths of these plateaus are rather size independent, see Table~\ref{tab:plateau_widths}; (ii) the exact magnetization jump close to saturation (above $m=7/9$) is very well reproduced; and (iii) we observe larger (respectively smaller) jumps above (respectively below) each plateau, which could indicate the absence (respectively presence) of a neighboring supersolid phase. 



\subsection{Nature of the magnetization plateaus at $m=1/3$, $5/9$, and $7/9$}\label{sec:high_field_plateaus}

As presented in detail below, we find that the states at the magnetization plateaus $m=\frac{1}{3}, \frac{5}{9}, \frac{7}{9}$ all spontaneously break lattice translation symmetry, forming $\sqrt{ 3 } \times \sqrt{ 3 }$ VBC states with an extended unit cell comprising nine lattice sites in the form of a hexagram, see Fig.~\ref{fig:observables_high_field_plateaus}, in agreement with earlier studies~\cite{Nishimoto2013,Capponi2013}. In analyzing local observables, we find that the hexagram unit cell naturally partitions into two inequivalent classes of sites, the hexagon sites that form the hexagram's inner hexagonal plaquette and the vertex sites located at its outer tips. We report for reference the average values (and fluctuations) taken by site magnetization and bond energies for each of these plateaus in Tables~\ref{tab:magnetization} and \ref{tab:bond_energies} of Appendix~\ref{sec:App-local}.

To go beyond the visual estimates of Fig.~\ref{fig:observables_high_field_plateaus}, we perform a symmetry-resolved characterization of the optimized plateau wave functions by decomposing them into irreps of the lattice's full space group $G$. To this end, we utilize the approach described in Sec.~\ref{sec:sym}. The stabilizer-based prediction for the irrep content of the $\sqrt{3} \times \sqrt{3}$ VBC is the same for all three high-field plateaus, as summarized in Table \ref{tab:irrep_content_both} and discussed below. The corresponding rescaled weights Eq.~\eqref{eq:rescaled_weights} obtained by Monte Carlo sampling are shown in Fig.~\ref{fig:irreps_13_59_79}, which will be discussed in the following description of all plateau states.

\subsubsection{$m=7/9$} \label{sec:7}
At $m=\frac{7}{9}$ the NQS ansatz converges to the  exact magnon crystal state with one  magnon per resonating hexagon~\cite{Schulenburg2002}. The sites on the vertices of the hexagram are fully polarized in the direction of the field, $m_{i}^{z}\approx\frac{1}{2}$, while the hexagon sites carry $m_{i}^{z}\approx\frac{1}{3}$, compatible with one delocalized magnon per plaquette. Thus, the average bond energy in each hexagon is $-1/12$, while it is equal to $1/6$ between the vertex and hexagon sites. 
As a result, the energy per site at zero field is found to be $\,e^{7/9}_{\text{NQS}}=0.1666(7)$, in agreement with the exact value $e^{7/9}_{\text{exact}}=1/6$.

As illustrated in Fig. \ref{fig:irreps_13_59_79}, the converged $m=7/9$ NQS has support only on the $\Gamma$ (either $\Gamma A1$ or $\Gamma B1$) and $K$ ($KA1$) sectors, as expected for a threefold degenerate magnon crystal state with $\sqrt{3}\times\sqrt{3}$ order. The observed switching between $\Gamma A1$ and $\Gamma B1$ sectors for even and odd $L$, respectively, is a commensurability effect which follows naturally from the structure of the magnon crystal as a product of one-magnon states with $k=\pi$ momentum confined on nonoverlapping hexagons.  A $\pi/3$ rotation flips the sign of the single-hexagon wave function, which results  in the full product state acquiring a factor $(-1)^{N_h}$, with $N_h=L^{2}/3$ being the number of resonating hexagons in the sample.  Hence, clusters with odd $N_h$ fall into $\Gamma B1$ rather than $\Gamma A1$, as observed for  the case $L=9$.

\subsubsection{$m=5/9$} \label{sec:5}
The same hexagram VBC pattern persists on the $m=\frac{5}{9}$ plateau [see Fig.~\ref{fig:triple:5_9_observables}], consistent with earlier studies
\cite{Nishimoto2013,Capponi2013}, however, with reduced local magnetizations. We find that the vertices are not fully polarized, see Table~\ref{tab:magnetization}, as expected, since the naive magnon crystal state is not an exact eigenstate anymore. The best variational
energy per site we obtain at this plateau is $e_{\text{NQS}}^{5/9}=-0.13500(1)$ for $L=6$ and $e_{\text{NQS}}^{5/9}=-0.13519(1)$ for $L=9$.

\begin{figure}[t]
    \centering
    \subfloat[]{
        \includegraphics[width=\columnwidth]{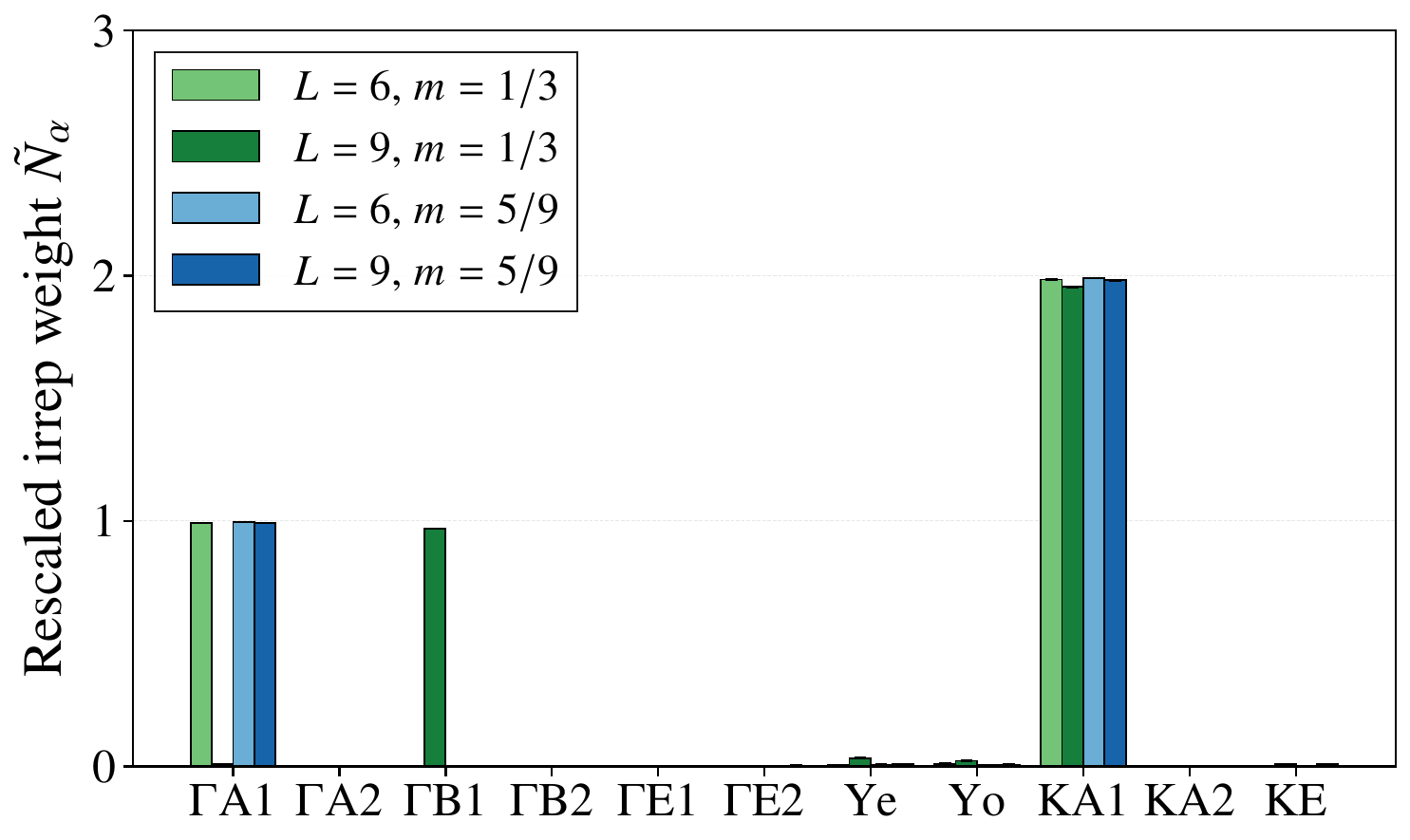}
        \label{fig:irreps_13_59}}\\[2mm]
    \subfloat[]{
        \includegraphics[width=\columnwidth]{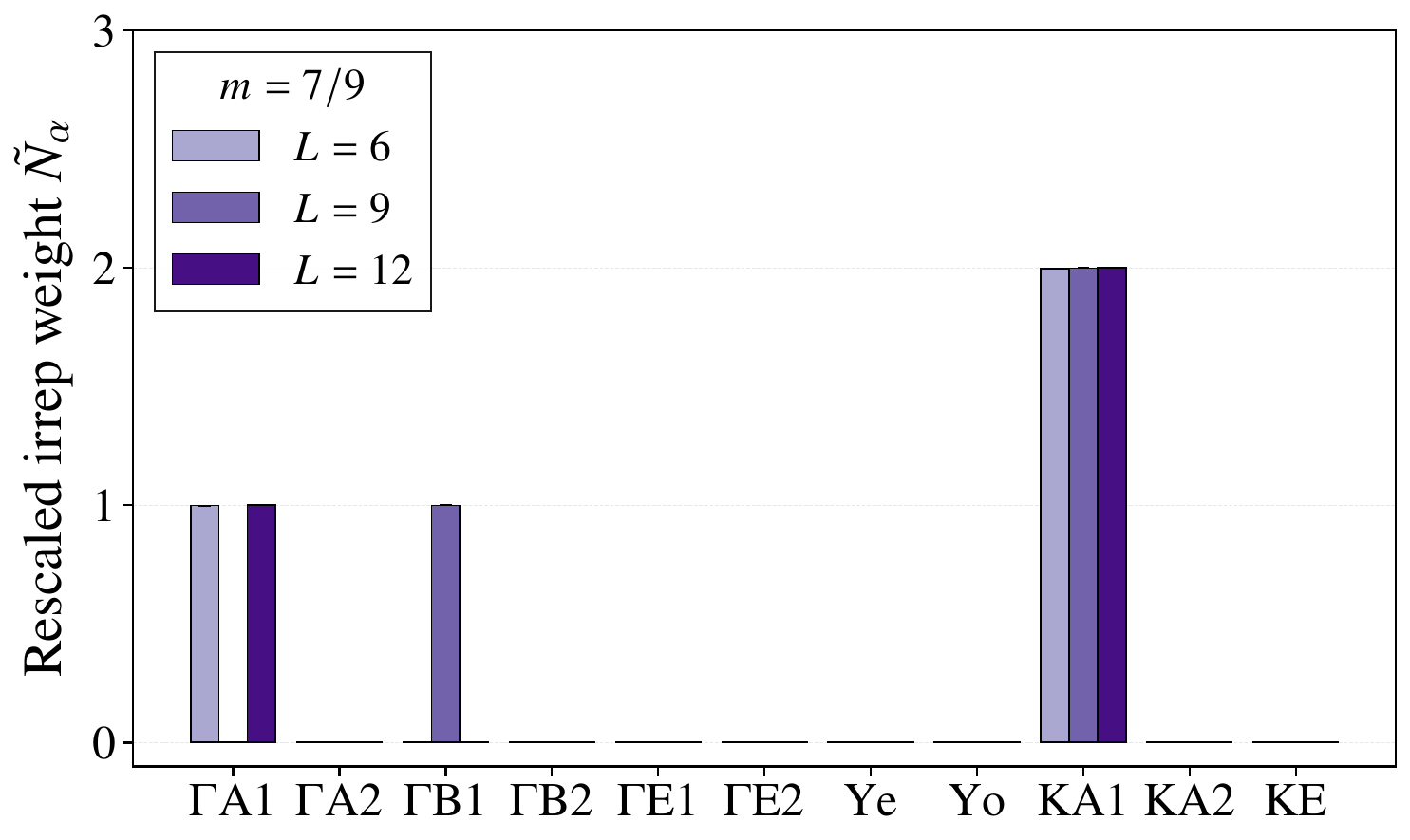}
        \label{fig:irreps_79}}
    \caption{Rescaled space group irrep weights $\tilde{N}_{\alpha}$ (see Eq. (\ref{eq:rescaled_weights})) for the optimized NQS plateau states at $m=1/3$, $5/9$, and $7/9$, for several system sizes. In all cases, the weight is concentrated in the $\Gamma$ and $K$ momentum sectors, consistent with the threefold-degenerate $\sqrt{3} \times \sqrt{3}$ VBC picture. The size-dependent $\Gamma A1$--$\Gamma B1$ irrep flipping for the $m=1/3$ and $7/9$ states can be attributed to the internal symmetry of the corresponding VBCs, see Sec.~\ref{sec:high_field_plateaus}.  The small residual leakage into other irreps at $m=1/3$ is a result of weak spatial fluctuations of the converged NQS (see Tables \ref{tab:magnetization} and \ref{tab:bond_energies} in Appendix~\ref{sec:App-local}). Irrep sectors not shown carry zero weight up to machine precision.}
    \label{fig:irreps_13_59_79}
\end{figure}

The optimized $m=5/9$ NQS retains the full point group symmetry of the kagome lattice, with no additional breaking of rotational symmetry as reported in iPEPS calculations \cite{Picot2016}.  
Similarly to the $m=7/9$ state, the irrep decomposition is fully compatible with a threefold degenerate $\sqrt{3}\times\sqrt{3}$ VBC. As shown in Fig. \ref{fig:irreps_13_59_79}, we find that, after 
projection into the space group irreps, the corresponding weights distribute as $w_{\Gamma}\approx 1/3$ in the $\Gamma$ sector and $w_K \approx 2/3$ in the twofold-degenerate $K$ sector.

Notably, the zero-momentum irrep remains 
$\Gamma A1$ for both $L=6$ and $L=9$, in contrast to the even-odd $\Gamma A1$ -- $\Gamma B1$ switching observed for $m=7/9$.  This behavior can be understood at the level of a hexagon-localized two-magnon VBC picture  \cite{Capponi2013}, which predicts a fully rotation-symmetric wave function on each resonating hexagon.

\subsubsection{$m=1/3$}\label{sec:3}
On the rather wide $m=1/3$ plateau that we have measured, local quantities shown in Fig.~\ref{fig:triple:1_3_observables} are again in agreement with the
same $\sqrt{ 3 } \times \sqrt{ 3 }$ VBC state, as found previously by other numerical techniques~\cite{Cabra2005,Nishimoto2013,Capponi2013,Picot2016,Chen2018}.
Similar to the $m=5/9$ plateau, the vertices are not fully polarized but they have a rather large magnetization (approximately 0.43, see Table~\ref{tab:magnetization}), which contradicts a recent proposal of a negative value (magnetization $-1/3$) found in another variational approach~\cite{He_2025} (with a variational energy per site close to $-0.342$). The variational energy per site obtained for the NQS state is equal to $\,e^{1/3}_{\text{NQS}}=-0.34657(2)$ for $L=6$ and $\,e^{1/3}_{\text{NQS}}=-0.34662(1)$ for $L=9$.

In terms of symmetries, the $m=1/3$ state retains the one-third, two-third weight splitting between the $\Gamma$ and $K$ momentum irreps as shown in Fig.~\ref{fig:irreps_13_59_79}. We note that the zero-momentum irrep content becomes again size-dependent, with a  $\Gamma A1$ -- $\Gamma B1$ switching appearing between even and odd $L$, similar to the $m=7/9$ magnon crystal state. This is in agreement with a much simpler ansatz variational state where each resonating hexagon would have $S_z^{\mathrm{hexagon}}=0$ and momentum $k=\pi$, i.e., its wave function changes sign under $2\pi/6$ rotation~\cite{Capponi2013}. Note that this ansatz wave function has a much higher variational energy~\cite{Capponi2013}: $e^{1/3}_\mathrm{Ansatz}\simeq  -0.311$.

\subsection{Nature of the state at $m=1/9$} \label{sec:9}

\begin{figure*}[!t] 
  \centering
  \subfloat[\large \vspace{-6mm} \textbf{VBC A}]{
    \includegraphics[
      width=0.45\textwidth,
      trim=45mm 20mm 45mm 20mm, 
      clip
    ]{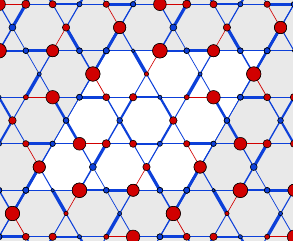}
    \label{fig:triple:VBC_A}
  }
  \hfill
  \subfloat[\large \vspace{-6mm} \textbf{VBC B}]{
    \includegraphics[
      width=0.45\textwidth,
      trim=45mm 20mm 45mm 20mm, 
      clip
    ]{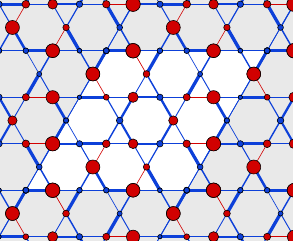}
    \label{fig:triple_VBC_B}
  }
  \vspace{5mm}
  \caption{Magnetization per site and bond energy   (see Eq.~\eqref{eq:mag_bond_energy})  for the VBC A and VBC B states obtained with unguided optimization for $L=6$ and $L=9$, respectively,  at the $m=1/9$ plateau. Disk diameter and bond width encode the magnitude of the observables, while red (blue) indicates a positive (negative) sign (see Appendix~\ref{sec:App-local} for numerical values). The unshaded region indicates the $3\times 3$ extended unit cell of the two VBCs.}
  \label{fig:observables_1_9}
\end{figure*}

We now turn to the more involved case, namely the plateau found at $m=1/9$.
Theoretically, the ground state at this magnetization has been proposed to be a topological gapped $\mathbb{Z}_3$ QSL~\cite{Nishimoto2013}, a gapped VBC (with an 18-fold degeneracy and a $\sqrt{3}\times\sqrt{3}$ structure~\cite{Picot2016, Morita2024}, or with 36-fold degeneracy and a $3\times 3$ structure~\cite{Cheng2026}), or a gapless $\sqrt{3}\times\sqrt{3}$ VBC~\cite{Fang2023}. Additional support for its $\mathbb{Z}_3$ QSL nature was provided by a recent VMC study~\cite{He2024_one_nine_Z3_QSL}, which would correspond to a gapped chiral topological QSL. At the experimental level, the $1/9$ plateau was first observed in Y-based kagome materials~\cite{Jeon2024,Suetsugu2024}. Further specific heat measurements point towards the possible signature of charge-neutral Dirac fermions~\cite{Zheng2025}. Additional evidence for low-energy fermionic excitations was found from unconventional magnetic oscillations~\cite{Zheng2025b}. All these results point to an exotic phase of matter.

We now turn to the results of our variational NQS investigation of the $m=1/9$ plateau. After extensive hyperparameter tuning and optimization, our results indicate that the state at the $1/9$ plateau breaks lattice translation symmetry and is characterized by a 27-site enlarged unit cell corresponding to $3\times 3$ order. We observe that the lowest-energy NQS for the $L=6$ and $L=9$ samples display different VBC patterns, as illustrated in the local observables of Fig. \ref{fig:observables_1_9}: we term VBC A (or `windmill'), the state that is found to minimize the energy for $L=6$ and VBC B, the one found for $L=9$. Their corresponding variational energies at zero field are  $E_A^{1/9}=-45.435(2)$ for $L=6$ [per site: $e^{1/9}_A=-0.42069(2)$]  and $E^{1/9}_B=-102.358(6)$ for $L=9$ [per site: $e^{1/9}_{B}=-0.42122(2)$].

We find that the two VBC patterns share a similar motif on their central hexagram, with a spontaneous symmetry breaking of the rotation. Indeed, the local magnetization on the hexagon is staggered, with three positive- and three negative values alternating around the ring, while all hexagon bonds have the same (negative) energy. The six vertices of the hexagram are all positively magnetized and connected to the hexagon with weak negative bonds. The difference between the two VBCs appears in the two bonds extending outwards from these vertices. Specifically, each vertex site has one outward bond with strongly negative energy while the other one is very weakly positive. In VBC A, these bonds appear in the same order as one moves clockwise from vertex to vertex, giving rise to a windmill-type motif, while in VBC B the order alternates between vertices. Consequently, the latter retains mirror symmetry along three of the six hexagram mirror axes, resulting in 18-fold degeneracy for VBC B in contrast to the 36-fold degeneracy of VBC A.

Comparing with previously proposed candidate states for the $1/9$ plateau, we observe that the VBC A and VBC B states achieve significantly lower variational energies per site than the $\mathbb{Z}_3$ QSL candidate, for which VMC   calculations yield $e=-0.41178$ \cite{He2024_one_nine_Z3_QSL} and the $\sqrt{3} \times \sqrt{3}$ VBC obtained with iPEPS \cite{Fang2023}, which gives $e=-0.4111$. We additionally observe that the VBC A  pattern closely resembles the windmill-type VBC  with $3 \times 3$ order found in Ref. \cite{Cheng2026}, which has variational energy per site equal to $-0.4184$. Beyond the  difference in variational energy, we note that the two states mainly differ in that VBC A hosts bonds with weakly positive-energy on the hexagram vertices, while all bonds of  the windmill VBC of Ref. \cite{Cheng2026} have negative energy. 
On the other hand,  the VBC B state has, to our knowledge,  not been reported previously in the literature and is distinguished by its higher point-group symmetry.

We note that obtaining the $1/9$ plateau state proved notably more challenging than the higher-field plateaus, with the optimizer exhibiting a tendency to become trapped in local minima with per-site energies in the range $e \in \left[ -0.4200, -0.4190\right]$.  These metastable states nevertheless consistently broke lattice translation symmetry up to the $3\times 3$ unit cell, never converging to patterns with smaller unit cells. We emphasize that both VBC A and VBC B emerged from unguided optimization for the $L=6$ and $L=9$ samples, subject only to the patch-translation invariance imposed by the NQS architecture. 

To further test whether the appearance of the VBC A and VBC B states on the two different samples is  due to size-dependent competition or an artifact of initialization or insufficient optimization, we  performed additional runs where each of the two VBC patterns was imprinted on a cluster of the alternate size (see Appendix~\ref{sec:App-imprinting} for details), namely, we imprinted an initial VBC B pattern for $L=6$, and an initial VBC A pattern for $L=9$. We then perform energy minimization starting from these initial states. In both cases, these runs converged to energies higher  than those obtained from unbiased training on the cluster of the same size (see Table~\ref{tab:energies_1_9} in Appendix~\ref{sec:App-optimization}), and moreover conserved the same symmetries as the ones initially imprinted. This indicates that the VBC A and VBC B patterns are indeed the more energetically stable states favored for $L=6$ and $L=9$, respectively.

\begin{figure}[t]
    \centering
    \subfloat[]{
        \includegraphics[width=\columnwidth]{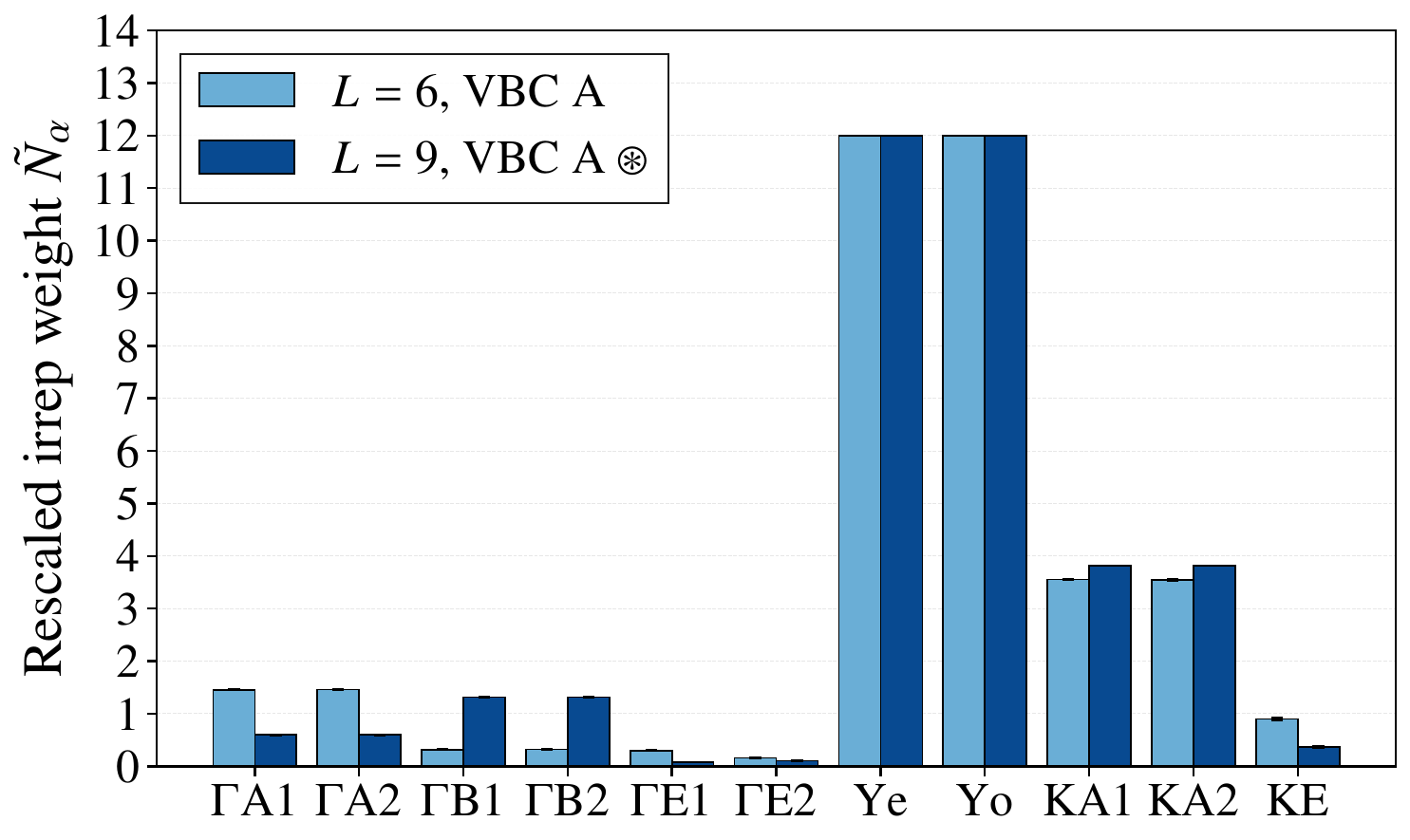}
        \label{fig:irreps_VBC_A}}\\[2mm]
    \subfloat[]{
        \includegraphics[width=\columnwidth]{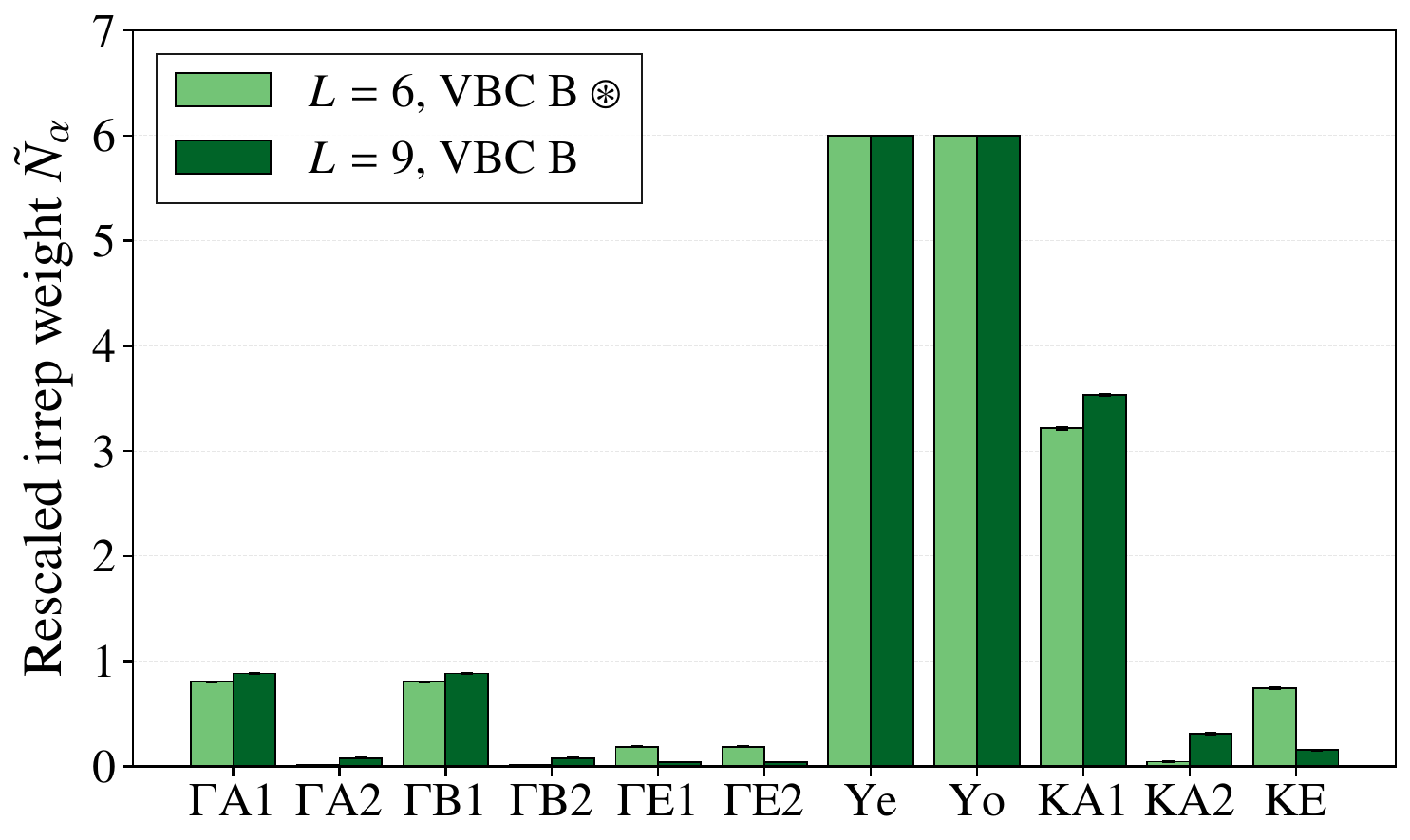}
        \label{fig:irreps_VBC_B}}
    \caption{Rescaled space group irrep weights $\tilde{N}_{\alpha}$ (see Eq. (\ref{eq:rescaled_weights})) for the optimized NQS plateau states at the $m=1/9$ plateau. The degeneracies used for rescaling are  $D_A=36$ for the VBC A and $D_B=18$ for the VBC B. Irrep sectors not shown carry zero weight up to machine precision, enforced by the patch-translation invariance of the ViT.  The VBC states obtained with imprinting are indicated with an asterisk.}
    \label{fig:irreps_VBCA_VBCB}
\end{figure}

To characterize more precisely the VBC patterns encountered in our analysis, we investigate the space-group irrep content of the low-energy optimized states, obtained from both unguided optimization or after imprinting the "wrong" VBC pattern. We use the two complementary symmetry decomposition methods described in Sec. \ref{sec:sym}.  
First, as a point of comparison, we computed the expected multiplicities $n_{\alpha}$, irrep dimensions $d_{\alpha}$, and total sector dimensions $N_{\alpha}$ for the two idealized versions of VBC A and VBC B, with results summarized in Table \ref{tab:irrep_content_both}. Next, we present in Fig. \ref{fig:irreps_VBCA_VBCB} the weights of each optimized state in each irrep by Monte Carlo estimation of the corresponding projector expectation value [Eq.~(\ref{eq:irrep_projector})].

Comparing the group-theoretic prediction with the weight decomposition  for the VBC A, we find the two  methods to be in qualitative agreement in the $Y$ and $K$ sectors for both the unguided ($L=6$) and the imprinted ($L=9$) optimization runs, up to some discrepancy attributable to weak spatial inhomogeneities of the converged NQS. We additionally observe that, while the character-stabilizer formula (\ref{eq:character_stabilizer}) predicts equal multiplicities in the four one-dimensional $\Gamma$ irreps, the numerical decomposition (\ref{eq:irrep_weight}) yields stronger support in the $A$-type irreps for the $L=6$ sample and $B$-type irreps for $L=9$. We attribute this to a finite-size effect owing to non-zero overlap between rotation-symmetry-related copies of the VBC A at small system sizes (see Appendix \ref{sec:App-irreps-A} for more details), with the stabilizer-based prediction being the favored decomposition in the thermodynamic limit. Both  methods are nevertheless in full agreement on predicting a mirror-symmetry-breaking state with $3\times 3$ order, notably distinct from the VBC B state.

Turning to the VBC B, we note that, on the $L=9$ sample, the rescaled projection weights $\tilde{N}_{\alpha}$ are in good agreement with the total irrep dimensions $N_{\alpha}$, confirming that the optimized NQS belongs to the expected symmetry-broken manifold.  The remaining discrepancies, most visible in the $KA1$ sector, can be attributed to a small leakage of weight into irreps that are absent in the ideal VBC pattern, again consistent with weak spatial inhomogeneities of the converged NQS. In contrast, the imprinted VBC B state on the $L=6$ sample exhibits a larger discrepancy, in line with the observation that it is a metastable local minimum with higher variational energy. Nevertheless, it retains qualitative agreement with the stabilizer-based prediction. Taken together with the aforementioned VBC A decomposition, this analysis thus indicates that the two competing VBC orders can be cleanly distinguished by their space-group irrep contents.

\begin{table}[t]
\centering
\newcommand{\VP}[2]{\begin{tabular}{@{}c@{\quad\,}c@{}}#1 & #2\end{tabular}}
\newcommand{\VPb}[2]{\begin{tabular}{@{}c@{\,\,\,\,}c@{}}#1 & #2\end{tabular}}
\begin{tabular}{l @{\hskip 1.0em} c @{\hskip 1.5em} ccc}
\toprule
 & & \textbf{VBC A} & \textbf{VBC B} & $\bm{\sqrt{3}\times\sqrt{3}}$ \textbf{VBCs} \\
\cmidrule(lr){3-3}\cmidrule(lr){4-4}\cmidrule(lr){5-5}
\textbf{Irrep} & $d_\alpha$
 & $n_\alpha$\,\,$N_\alpha$
 & $n_\alpha$\,\,$N_\alpha$
 & $n_\alpha$\,\,$N_\alpha$ \\
\midrule
$\Gamma A1$      & 1 & \VP{1}{1}  & \VP{1}{1} & \VP{1}{1} \\
$\Gamma A2$      & 1 & \VP{1}{1}  & \VP{0}{0} & \VP{0}{0} \\
$\Gamma B1$      & 1 & \VP{1}{1}  & \VP{1}{1} & \VP{0}{0} \\
$\Gamma B2$      & 1 & \VP{1}{1}  & \VP{0}{0} & \VP{0}{0} \\
$\mathrm{K}A1$   & 2 & \VP{2}{4}  & \VP{2}{4} & \VP{1}{2} \\
$\mathrm{K}A2$   & 2 & \VP{2}{4}  & \VP{0}{0} & \VP{0}{0} \\
$Y_e$            & 6 & \VPb{2}{12} & \VP{1}{6} & \VP{0}{0} \\
$Y_o$            & 6 & \VPb{2}{12} & \VP{1}{6} & \VP{0}{0} \\
\bottomrule
\end{tabular}
\caption{Irrep decomposition of the degenerate ground-state manifolds for all VBC states considered in our study, as predicted from the character-stabilizer formula of Eq. (\ref{eq:character_stabilizer}). We take the stabilizer to contain the residual translations in all cases, as well as point groups generated by $C_3$ rotation for the VBC A, $C_3$ rotation and reflection for the VBC B, while we consider full $C_{6v}$ point-group symmetry for the  $\sqrt{3}\times\sqrt{3}$ high-field VBC states,  for which the formula yields the same irrep content. Let us remind that Eq.~(\ref{eq:character_stabilizer}) provides a simplified counting rule which does not take into account any nontrivial internal structure of the representative state. See Appendices \ref{sec:App-symmetry} and \ref{sec:App-irreps-A} for more details.} 
\label{tab:irrep_content_both}
\end{table}

\section{Discussion and conclusions}\label{sec:conc}

Our results for the magnetization plateaus $m=7/9,5/9,1/3$ confirm that they host valence bond crystalline ground states, as was discussed previously. For the $m=1/3$ plateau, ViT NQS provide the lowest variational energy reported so far (to the best of our knowledge), reaching  $e^{m=1/3}=-0.34657(2)$ and $e^{m=1/3}=-0.34662(1)$ for $L=6$ and $L=9$, respectively. The magnetization and bond-energy patterns observed for $m=1/3$ (see Tables~\ref{tab:magnetization} and~\ref{tab:bond_energies}) are in close agreement with the recent work~\cite{Cheng2026}, which used fermionic variational wave functions, but strongly differ from the ones proposed in Ref.~\cite{He_2025} which finds negative ({\it i.e.} field-opposed) magnetization for spins outside of hexagons.

For the $m=1/9$ plateau, ViT NQS also appear to obtain the lowest variational energies reported so far (see note below however), and conclude in favor of a Valence Bond Crystal with a large 27-site cell, within the limitations of our study discussed below. The optimization of ViT NQS for $m=1/9$ is challenging, suggesting that there are many local energy minima within the explored variational landscape. This is in line with the numerous types of low-energy states of different natures found in other variational studies ranging from a topological $\mathbb{Z}_3$ QSL ~\cite{Nishimoto2013, He2024_one_nine_Z3_QSL} to several candidate VBCs with distinct lattice symmetry-breaking and enlarged unit cells~\cite{Picot2016,Fang2023,Morita2024, Cheng2026}, most having comparable energy per site, e.g., $-0.4111$~\cite{Fang2023} or $-0.4184$~\cite{Cheng2026}. The VBC we find for the $L=6$ ($N=108$) sample agrees with the one obtained in Ref.~\cite{Cheng2026} ('windmill'), but we find that a slightly different, higher-symmetrical, VBC pattern (VBC B) is more stable on the $L=9$ sample. Our stability analysis, conducted by initializing runs with competing VBC patterns, confirms these results. We find that unguided optimization consistently yields lower energies for a comparable number of iterations. Furthermore, the persistence of the imprinted pattern's symmetry—despite having a variational energy only slightly higher than the ground state, indicates the existence of competing states within a narrow low-energy window for $m=1/9$ at these values of $L$.

We now briefly mention the limitations of our approach. First, it is important to recall that NQS is a variational method and thus not guaranteed to find the exact ground state. However, the fact that our energies for the $m = 1/3, 5/9, \text{ and } 7/9$ plateaus match or improve upon previous studies confirms that our approach accurately captures the underlying VBC physics. Next, the choice of the patch (with 27 sites) together with the patch translation invariance does not allow us to capture ground states with a larger unit cell. This can be remedied with the choice of a larger patch, albeit at a very high computational cost. Also, the optimization landscape for the $m=1/9$ plateau appears complex, with many competing candidate states, and it is not guaranteed that we reached the lowest energy minimum despite our thorough optimization process. In Appendix~\ref{sec:App-protocol}, we present details of the energy minimization process, where we conclude that while a slightly lower energy could perhaps be obtained by performing longer runs, the physical nature of the states found would not be affected. Finally, our results for the $m=1/9$ VBC ground states which are not exactly the same for the $L=6$ and $L=9$ samples do not allow us to conclude on the nature of the ground state for larger systems or in the thermodynamic limit. We have attempted long runs for the $L=12$ sample (with N=432 spins) using the same ViT NQS architecture, but find that the optimization persistently becomes trapped in initialization-dependent local minima for thousands of optimization steps. Combined with the high computational cost at this system size, the frequent restarts did not allow us to reach convergence in a reasonable computational time for this system size. We believe however that our results point towards a VBC ground state with a large (at least 27-sites) unit cell for the $m=1/9$ plateau. This prediction could be tested experimentally, e.g., by performing Nuclear Magnetic Resonance experiments on this plateau to probe the local magnetization values.

We finish by discussing other possible future applications of NQS to plateau physics in frustrated magnets, by first considering extensions for the kagome lattice. Exotic phases can also emerge at finite magnetization density when considering the XXZ anisotropy of the Heisenberg model. While most plateaus are expected to be robust~\cite{Kshetrimayum2016,Huerga2016}, a novel topological QSL phase has been predicted at $m=2/3$ in the XY limit~\cite{Kumar2014,Kumar2016}. Moreover, when considering extended interactions, a nematic plateau at $m=1/6$ has also been found~\cite{Plat2015}. The NQS approach can also capture supersolid phases, which are often found near plateaus, see, e.g., Refs.~\cite{MomoiTotsuka2000b,Plat2018,huang2026emergentspinsupersolidsfrustrated}, which would require an analysis of out-of-plane spin-correlations. Larger spin values could also stabilize unconventional plateaus~\cite{Picot2016}. 
Regarding spin-1/2 kagome materials, herbertsmithite seems to exhibit additional plateaus which could be stabilized by further neighbor interactions~\cite{Okuma2019}. Other materials, such as kapellasite, probably contain some additional deformations that could lead to other kinds of plateaus \cite{Okuma2019,Bodaiji2024}. 

Using the same methodology, we could also investigate similar magnetization plateaus found on other 2d frustrated lattices, e.g., checkerboard~\cite{Morita2016,Capponi2017} or Husimi~\cite{Liao2016} lattices. 
It would be also worth investigating finite-temperature properties, for instance to make contact with ongoing experiments and understand at which temperatures the plateaus could be observed and above which the crystal phases  melt and disappear~\cite{Schnack2018,Misawa2020,Schnack2020,Schlueter2022}. Quite interestingly, NQS approaches could be used as well~\cite{Lange2024}.

{\it Note added.} Recently, we became aware of Ref.~\cite{Duric2025b}, which uses a very similar variational NQS approach with group convolutional neural networks (GCNN) for the same $m=1/9$ kagome plateau. Reference~\cite{Duric2025b} finds a variational state with a lower energy that is interpreted as a gapless chiral spin density wave, quite different from the VBCs that we find. Given that the two architectures (GCNN and ViT with patch translation invariance, and symmetry projection on space-group irreps) and methods of optimization are very similar, we do not have a simple explanation for the sizable difference between our best variational energy per site (e.g., $e_0^{m=1/9}\simeq -0.4207 $  for the $L=6$ sample) versus the ones ($e_0^{m=1/9}\simeq -0.4984$) reported in Ref.~\cite{Duric2025b}. We note that the (zero-field) energy per site $e_0^{m=1/9}$ reported in Ref.~\cite{Duric2025b} for the magnetized state $m=1/9$ is {\it lower} than the best known values for the energy per site for a {\it nonmagnetic state} (e.g., $e_0^{m=0}=-0.4461$ in Ref.~\cite{Duric2025a}, or several numerical studies pointing to a ground-state energy per site $\in [-0.4384,-0.42866]$, see Ref.~\cite{zhu_quantum_2025}), which is not expected. We further derive in Appendix~\ref{app:bound} an exact lower bound for the zero-field energy $e_0^{m=0}\geq -0.4752$, which appears to contradict the quoted values of Ref.~\cite{Duric2025b}.

\begin{acknowledgments}
We acknowledge useful discussions with G. Misguich, R. Nutakki, D. Poilblanc, A. Szab\'o, F. Vicentini and A. Wietek. 
We thank CALMIP (Grant No. 2025-P0677 on Olympe cluster, Project No. M25154 on Turpan cluster) and GENCI (Project No. A0190500225) for computer resources. This work benefited from the support of the project QMAHT, Grant No. ANR-22-CE30-0032-03, of the French National Research Agency (ANR). 
Our NQS simulations were performed using NetKet~\cite{netket3_2022, netket2_2019} which is built on top of JAX~\cite{jax2018github}, Flax~\cite{flax2020github}, and Optax~\cite{hessel2020optax} libraries.

\end{acknowledgments}

\appendix

\section{Statistics of local observables}
\label{sec:App-local}
To provide a quantitative characterization of the VBC states observed at the magnetization plateaus discussed in the main text, we report statistics of the corresponding local magnetizations and bond energies, defined as 

\begin{equation}\label{eq:mag_bond_energy}
m_i^{z}\equiv \langle S_i^{z}\rangle,\qquad \varepsilon_{ij}\equiv \langle \mathbf S_i\!\cdot\!\mathbf S_j\rangle .
\end{equation}
We compute site- and bond-averaged values
\begin{equation}
m^{z}_{I}= \frac{1}{|I|}\sum_{i\in I} m_i^{z},\qquad
\varepsilon_{B}= \frac{1}{|B|}\sum_{\langle i,j\rangle\in B} \varepsilon_{ij},
\end{equation}
together with the corresponding spatial standard deviations
\begin{equation}
\begin{aligned}
\sigma_{m,I}&=\left[\frac{1}{|I|}\sum_{i\in I}\big(m_i^{z}-m^{z}_{I}\big)^{2}\right]^{1/2},\\
\sigma_{\varepsilon,B}&=\left[\frac{1}{|B|}\sum_{\langle i,j\rangle\in B}\big(\varepsilon_{ij}-\varepsilon_{B}\big)^{2}\right]^{1/2}.
\end{aligned}
\end{equation}
Here $I$ denotes the chosen set of sites (e.g., hexagon or vertex sites in the case of the $\sqrt{3} \times \sqrt{3}$ VBCs) and $B$ the set  of selected nearest-neighbor bonds.

\begin{figure}[t]
    \centering
    \subfloat[]{
        \includegraphics[width=\columnwidth]{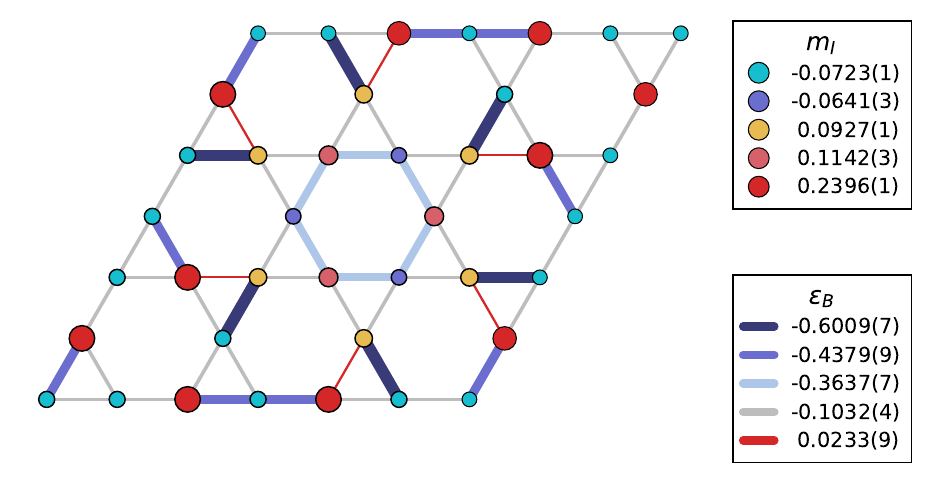}
        \label{fig:vbc_a_detail}}\\[0.2mm]
    \subfloat[]{
        \includegraphics[width=\columnwidth]{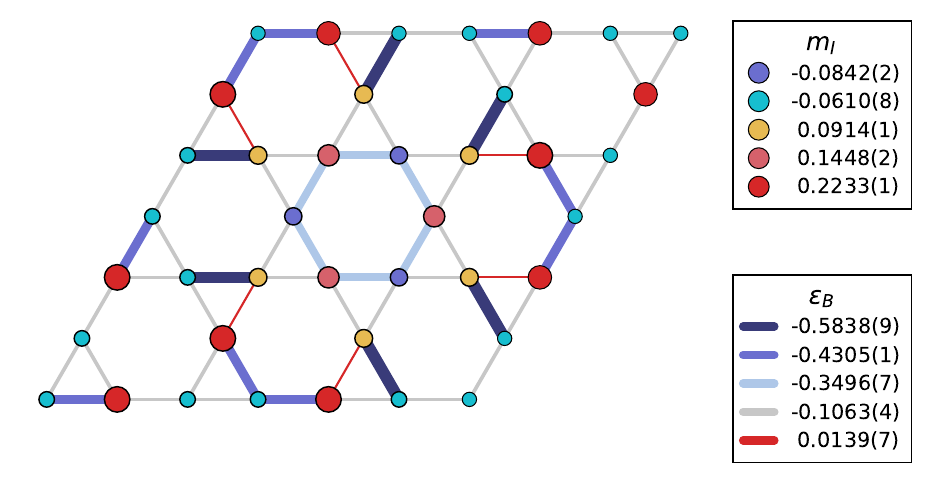}
        \label{fig:vbc_b_detail}}
    \caption{Spatially averaged values for the magnetization and bond energy for the VBC A (top) and VBC B (bottom) states obtained with unguided optimization at the $m=1/9$ plateau (see Sec.~\ref{sec:9}). Averaging is carried out over elements of the same color. The radius of the circles and width of the bonds illustrate the magnitude of the data.}
    \label{fig:1_9_observables_detailed}
\end{figure}

The resulting data for the three high-field plateaus are summarized in Tables \ref{tab:magnetization} and \ref{tab:bond_energies}, while the corresponding mean values for the VBC A and VBC B states obtained with unguided optimization at the $m=1/9$ plateau are illustrated in Fig.~\ref{fig:1_9_observables_detailed}.

\begin{table}
  \renewcommand{\arraystretch}{1.17}
  \caption{\label{tab:magnetization} Local magnetization statistics for the VBC states obtained at the $1/3$, $5/9$, and $7/9$ plateaus. We report the site-set expectation value $m_I^z$ and spatial standard deviation $\sigma_{m,I}$ for the vertex and hexagon sites, obtained with $M=2^{16}$ MC samples.}
  \begin{ruledtabular}
    \begin{tabular}{lccc}
      Plateau & Site set &  $m_I^z$ &  $\sigma_{m,I}$ \\ \colrule
      $1/3$ & Vertex  & $0.4273(2)$ & $0.001$ \\
      $1/3$ & Hexagon & $0.0364(1)$ & $0.015$ \\
      $5/9$ & Vertex  & $0.47264(5)$ & $0.001$ \\
      $5/9$ & Hexagon & $0.18034(5)$ & $0.003$ \\
      $7/9$ & Vertex  & $0.4999971(7)$    & \phantom{0.00}$0$ \\
      $7/9$ & Hexagon & $0.3333347(3)$    & $0.001$ \\
    \end{tabular}
  \end{ruledtabular}
\end{table}


\begin{table}
  \renewcommand{\arraystretch}{1.18}
  \caption{\label{tab:bond_energies} Bond-energy statistics for the VBC states obtained at the $1/3$, $5/9$, and $7/9$ plateaus. We report the bond-set expectation value $\varepsilon_B$ and spatial standard deviation $\sigma_{\varepsilon,B}$ for the indicated nearest-neighbor bond sets, obtained with $M=2^{16}$ MC samples.}
  \begin{ruledtabular}
    \begin{tabular}{lccc}
      Plateau & Bond set &  $\varepsilon_B$ &  $\sigma_{\varepsilon,B}$ \\ \colrule
      $1/3$ & Hexagon--Vertex  & $-0.0603(1)$ & $0.013$ \\
      $1/3$ & Hexagon--Hexagon & $-0.3993(1)$ & $0.002$ \\
      $5/9$ & Hexagon--Vertex  & $\phantom{-}0.05861(5)$ & $0.003$ \\
      $5/9$ & Hexagon--Hexagon & $-0.31996(9)$ & $0.004$ \\
      $7/9$ & Hexagon--Vertex  & $\phantom{-}0.1666643(5)$ & $0.001$ \\
      $7/9$ & Hexagon--Hexagon & $-0.083328(1)$ & $0.002$ \\
    \end{tabular}
  \end{ruledtabular}
\end{table}

\section{Space-group irrep labeling}
\label{sec:App-symmetry}

\begin{figure}
    \centering
    \includegraphics[width=\columnwidth]{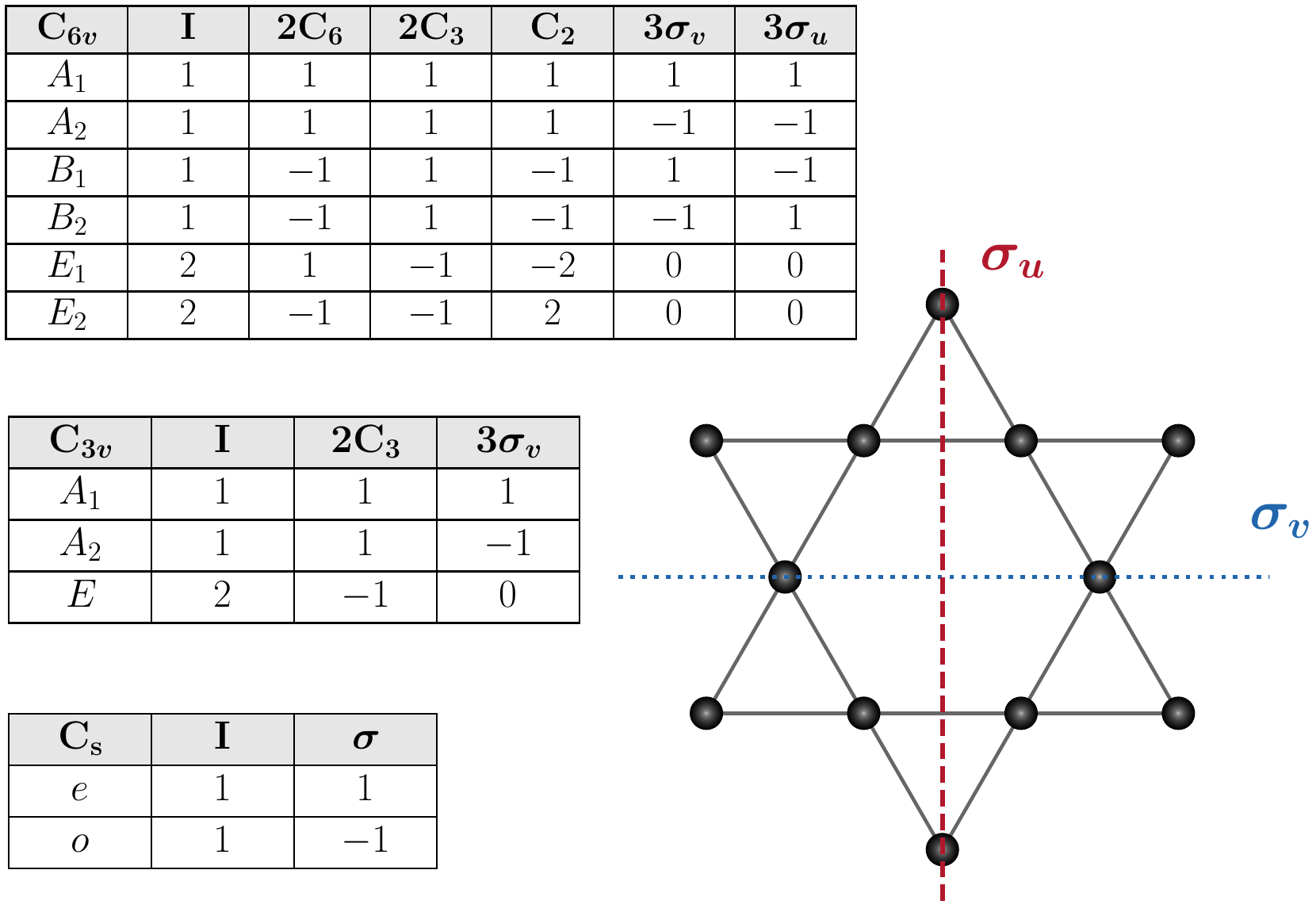}
    \caption{Group elements and character tables for the little cogroups of the momentum stars captured by our choice of ViT patch (see Fig.~\ref{fig:lattice_and_bz}).}
    \label{fig:kagome_symmetries}
\end{figure}

For a kagome lattice sample with $L\times L$ primitive unit cells and periodic boundary conditions, the corresponding  space group is the semidirect product 

\begin{equation}
    G = T \rtimes C_{6v},
\end{equation}
where $T\simeq \mathbb{Z}_L \times \mathbb{Z}_L$ is the finite translation group generated by Bravais translations $T_{\mathbf{a_1}}$ and $T_{\mathbf{a_2}}$, while $C_{6v}$ is the order-12 group generated by a  $C_6$ ($2\pi/6$) rotation and a reflection $\sigma$ about a line through the rotation axis (see Fig.~\ref{fig:kagome_symmetries}). 

To classify the irreps of the space group $G$, we define the star $[\mathbf{k}]$, the set of distinct Brillouin zone momenta obtained from the action of the lattice point group on a chosen BZ momentum $\mathbf{k}$. Let us additionally denote with $G_{\mathbf{k}}$ the little cogroup for the momentum $\mathbf{k}$, that is, the subgroup  of point group symmetries that leave $\mathbf{k}$ invariant up to a reciprocal lattice vector. A space-group irrep is then labeled by a choice of momentum star $[\mathbf{k}]$ (hereby  simply denoted by the representative momentum $\mathbf{k}$) together with an irrep $\rho$ of $G_{\mathbf{k}}$. 

The  stars relevant for the present work are those associated with the elements of the folded set $\mathcal{K}_{\Gamma}$ [see Eq.~(\ref{eq:vit_momenta}) in the main text], that is, the momenta representable by our ViT with the patch choice described in Sec.~\ref{sec:sym}. They correspond to the three sets $[\Gamma]=\{\Gamma\}$, $[K]=\{K,K'\}$, and $[Y]=\{Y_i\}_{i=0}^5$. The associated little cogroups are 

\begin{equation}
    G_{\Gamma} \simeq C_{6v}, \quad G_K \simeq C_{3v}, \quad G_Y \simeq C_{s}
\end{equation}
where $C_{3v}$ is the six-element group generated by a $C_3$ ($2\pi/3$) rotation and a reflection,  and $C_{s}$ being an order-2 group generated by a single reflection. The resulting space-group irreps $\alpha$ can thus be labeled with the star representative and the little cogroup  irrep label. To specify our little cogroup labeling convention, we provide  the corresponding character tables along with a schematic defining our symmetry-element conventions in Fig.~\ref{fig:kagome_symmetries}.

\section{Symmetry analysis details for the VBC A state}\label{sec:App-irreps-A}

Based on the real-space pattern of local observables (Figs. \ref{fig:observables_1_9} and \ref{fig:1_9_observables_detailed}), the VBC A state can be interpreted as breaking the $C_{6v}$ kagome point group down to a subgroup generated solely by $C_3$ rotations, in addition to the translation symmetry breaking. Treating this idealized symmetry-broken pattern as corresponding to a representative state $\ket{\Phi_A}$, the character-stabilizer formula (\ref{eq:character_stabilizer}) yields the irrep content reported in Table~\ref{tab:irrep_content_both}, predicting equal multiplicities among the four one-dimensional $\Gamma$ irreps and 36-fold degeneracy.

On finite clusters, however, the variationally-optimized state $\ket{\Psi_A}$ may not coincide with an idealized representative $\ket{\Phi_A}$ and the distinct images of the VBC pattern under point group action may not be exactly orthogonal. To diagnose potential finite-size effects, we thus additionally estimate by Monte Carlo sampling the expectations $\langle \Psi_A | U(g) |\Psi_A \rangle$, corresponding to the overlap between symmetry-transformed VBC A states, for the point-group elements $g=C_6, C_3,\sigma_v, \sigma_u$. While all reflection expectations are zero to machine precision, the overlaps under rotations are found to be $\langle U(C_6)\rangle\simeq 0.57\, (L=6)$ and $\simeq -0.37\, (L=9)$, while $\langle U(C_3)\rangle\simeq 0.83\, (L=6)$ and $\simeq 0.94\, (L=9)$. These values indicate that, while the optimized finite-size VBC A states fully break reflection symmetry, they retain significant overlap under $C_6$ rotations, which nevertheless rapidly decreases with increasing system size. 

This observation allows for a direct interpretation  of the numerical irrep decomposition data shown in Fig. \ref{fig:irreps_VBCA_VBCB}. Augmenting the stabilizer for the (finite-size) VBC A state to also include $C_6$ rotations (while still excluding reflections), the character-stabilizer formula (\ref{eq:character_stabilizer}) predicts support only in the A-type $\Gamma$ irreps (see Table~\ref{tab:irreps_c6_vbc_a}), consistent with the $L=6$ numerical irrep decomposition data of Fig. \ref{fig:irreps_VBCA_VBCB} (after appropriate rescaling to account for the now-reduced 18-fold degeneracy). Moreover, the change of sign of $\langle U(C_6) \rangle$ between $L=6$ and $L=9$ also accounts for the $A-B$ exchange observed for the one-dimensional $\Gamma$ irreps between even and odd system sizes, respectively (see corresponding characters for the $2C_6$ conjugacy class in the $C_{6v}$ table of Fig. \ref{fig:kagome_symmetries}). 

\begin{table}[t]
    \centering
    {\renewcommand{\arraystretch}{1.2}
    \begin{tabular}{|c|c|c|c|c|c|c|}
    \hline
         & $\Gamma A1$ & $\Gamma A2$ & $Ye$ & $Yo$ & $K A1$ & $K A2$  \\
        \hline
      $N_\alpha$   & $1$ & $1$ & $6$ & $6$ & $2$ & $2$ \\
      \hline
    \end{tabular}}
    \caption{Irrep decomposition of an 18-fold degenerate ground-state manifold predicted from the character-stabilizer formula (\ref{eq:character_stabilizer}) for an idealized VBC state with a $3\times3$ extended unit cell, full $C_6$ rotation symmetry and reflection-symmetry breaking.  }
    \label{tab:irreps_c6_vbc_a}
\end{table}

Notably, the magnitude of the $C_6$ ($C_3$) expectation value decreases (increases) with system size, indicating a trend towards a state with only $C_3$ point group symmetry and consistent with the expectation that distinct symmetry-related VBC images should become orthogonal at the thermodynamic limit. This observation is also in agreement with the more uniform division of the projected weight in the four $\Gamma$ irreps observed in Fig.~\ref{fig:irreps_VBCA_VBCB} for $L=9$, relative to $L=6$.

We thus believe that the stabilizer-based counting of Table~\ref{tab:irrep_content_both} in the main text should be understood as the thermodynamic-limit decomposition of the degenerate ground-state manifold for the VBC A state, with the aforementioned finite-size effects accounting for the $\Gamma$-irrep data of Fig.~\ref{fig:irreps_VBCA_VBCB}. Nevertheless, this effect does not alter the primary conclusion that the state breaks translation symmetry  down to a $3\times 3$ extended unit cell  or that it exhibits robust reflection-symmetry breaking that distinguishes it from the VBC B state.

\section{Lower bound for the zero-field energy}
\label{app:bound}
Here we briefly describe how we derive the exact lower bound for the zero-field energy per site which is quoted in the Note added of Sec.~\ref{sec:conc}, using the method of Anderson bounds~\cite{Anderson1951,kránitz2026}. 

Given a Hamiltonian $H$ and a choice of decomposition $H = \sum_i H_i$ into sub-Hamiltonians $H_i$, the variational principle guarantees that, for all sub-Hamiltonians, we have $
  \langle \psi_0 | H_i | \psi_0 \rangle \geq E_0^{(i)}$, 
with $\ket{\psi_0}$ being a ground state of $H$ and $E_0^{(i)}$ the ground-state energy of $H_i$. Combining these equations, we obtain
\begin{equation}\label{eq:anderson_bound}
  E_{\mathrm{GS}}\geq \sum_i E_0^{(i)},
\end{equation}
which provides a lower bound for the ground-state energy $E_{\mathrm{GS}}=\langle \psi_0 | H | \psi_0 \rangle$ of $H$. 

Denoting by $H_L$ the Heisenberg Hamiltonian (\ref{eq:model}) on a kagome lattice sample with $L\times L$ primitive (three-site) unit cells and periodic boundary conditions, we choose the decomposition

\begin{equation}\label{eq:decomposition_cluster}
  H_L=\sum_{\boldsymbol{R}} H_{\boldsymbol{R}}^{\mathcal{C}}
\end{equation}
where the sum runs over all $L^2$ Bravais lattice translation vectors $\boldsymbol{R}$. Letting $T_{\boldsymbol{R}}$ denote the corresponding translation operator, the sub-Hamiltonians are defined as

\begin{equation}\label{eq:subhamiltonian_translations}
  H_{\boldsymbol{R}}^{\mathcal{C}}=T_{\boldsymbol{R}} H^{\mathcal{C}}T_{\boldsymbol{R}}^{\dagger},
\end{equation}
that is, by translation of a reference sub-Hamiltonian $H^{\mathcal{C}}$ defined on a cluster $\mathcal{C}$ of $N_{\mathcal{C}}$ sites with open boundary conditions. 

Since all sub-Hamiltonians $H_{\boldsymbol{R}}^{\mathcal{C}}$ are unitarily equivalent to $H^{\mathcal{C}}$, an Anderson bound for the ground-state energy $E_{GS}(L)$ of $H_L$ can be obtained directly from the smallest eigenvalue, $E_0^{\mathcal{C}}$, of the reference cluster Hamiltonian. The resulting inequality reads

\begin{equation}
    E_{GS}(L)\geq L^2 E_0^{\mathcal{C}}
\end{equation}
or, in terms of energy per site,

\begin{equation}\label{eq:anderson_inequality_per_site}
    e_{GS}(L) \geq \frac{L^2}{3L^2}E_0^{\mathcal{C}} = \frac{1}{3} E_0^{\mathcal{C}}.
\end{equation}
Since the right-hand side has no dependence on $L$, the quantity
\begin{equation}
    B^\mathcal{C}\equiv \frac{1}{3} E_0^{\mathcal{C}}
\end{equation}
provides a lower bound for the energy per site of the kagome Heisenberg Hamiltonian (\ref{eq:model}) at the thermodynamic limit.

\begin{figure}[t]
\includegraphics[width=0.5\columnwidth]{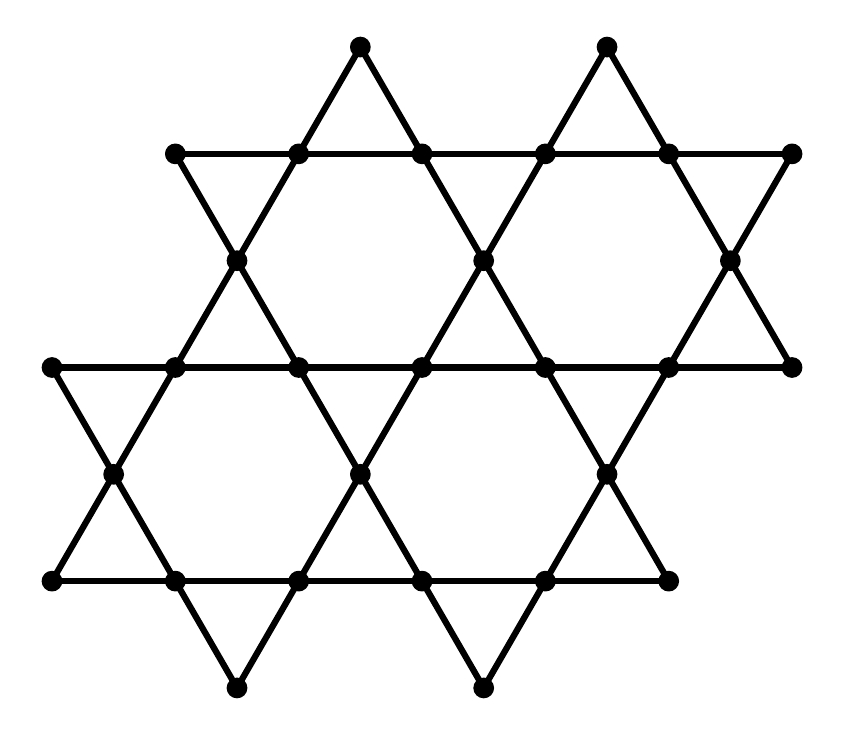}
\caption{The 29-site cluster $\mathcal{C}_{29}$ with open boundary conditions used to derive the bound (\ref{eq:bound_29}).} 
\label{fig:cluster_29}
\end{figure}

To obtain the bound quoted in the main text, we consider the reference sub-Hamiltonian

\begin{equation}\label{eq:hamiltonian_cluster_29}
H^{\mathcal{C}_{29}}= \frac{J}{8} \sum_{\langle i,j \rangle} \boldsymbol{S}_i\cdot \boldsymbol{S}_j,
\end{equation}
where the sum runs over nearest-neighbor sites on the 29-spin cluster with open boundary conditions shown in Fig. \ref{fig:cluster_29}. The uniform $J/8$ coupling  is chosen such that Eq. (\ref{eq:decomposition_cluster}) is satisfied.   

Using Exact Diagonalization, we obtain its ground-state energy $E_0^{\mathcal{C}_{29}}=-1.42558084$  
which yields the  bound 

\begin{equation}\label{eq:bound_29}
B^{\mathcal{C}_{29}}=-0.47519361
\end{equation}
for the zero-field energy per site. 

Since the zero-field ground-state energy in any fixed-magnetization sector is bounded from below by the unrestricted ground-state energy,  the Anderson-type bound derived above also applies to the zero-field energy density in the $m=1/9$ sector, namely,

\begin{equation}\label{eq:bound_1_9}
e_0^{m=1/9}\ge B^{\mathcal C_{29}}.
\end{equation}

\section{Optimization details for plateau states}
\label{sec:App-optimization}

In this Appendix, we describe the protocol used for optimization on the $m=1/9$, $1/3$, and $5/9$ magnetization plateaus as well as the imprinting procedure we used to bias the NQS training towards targeted VBC states at $m=1/9$, in order to compare with those obtained with unguided optimization (see Sec.~\ref{sec:9}).

\begin{table}[!b]
\centering
\setlength{\tabcolsep}{8pt}
\renewcommand{\arraystretch}{1.4}
\footnotesize
\begin{tabular}{lcc}
\toprule
 & $L=6$ & $L=9$ \\
\midrule
\addlinespace[6pt]
\makecell{Best energy \\ (unguided)} &
\makecell{VBC A\\ $e^{1/9}=-0.42069(2)$\\ $E^{1/9}=-45.435(2)$} &
\makecell{VBC B\\ $e^{1/9}=-0.42122(2)$\\ $E^{1/9}=-102.358(6)$} \\
\addlinespace[10pt]
\makecell{Imprinted} &
\makecell{VBC B\\ $e^{1/9}=-0.41899(2)$\\ $E^{1/9}=-45.251(2)$} &
\makecell{VBC A\\ $e^{1/9}=-0.42093(2)$\\ $E^{1/9}=-102.285(4)$} \\ 
\bottomrule
\end{tabular}
\caption{Variational zero-field energies per site ($e^{1/9}$) and total ($E^{1/9}$) for the two VBC states on the $m=1/9$ plateau. ``Imprinted" denotes optimization initialized from the chosen VBC pattern (see Appendix \ref{sec:App-imprinting}), while ``unguided" denotes unconstrained optimization starting from random initialization. Imprinted states appear to remain metastable.}
\label{tab:energies_1_9}
\end{table}

\subsection{Imprinting methodology for VBCs at $m=1/9$}\label{sec:App-imprinting}

To bias the optimization towards a targeted VBC, we start by training a ViT NQS under a modified Heisenberg Hamiltonian where the coupling on selected sets of bonds (different for each target state) is increased by $\delta J=0.2 J$, together with an additional staggered longitudinal field term with amplitude $h^\prime=0.2 J$ applied on the hexagon sites. For  VBC A, the modified bonds correspond to the  ``arms" of the windmill pattern together with the links forming the central hexagon of the $3 \times 3$ unit cell (see Fig. \ref{fig:observables_1_9}). In the VBC B case, we again modify the hexagon bond couplings along with the couplings on the links hosting the strong bonds emanating from the vertices of the hexagram (see Fig. \ref{fig:observables_1_9}). We note that, since Monte Carlo sampling is performed within a fixed $S^z$ sector,  the added staggered external field only serves to redistribute the local magnetization without changing its total value.

We optimize the NQS under this perturbed Hamiltonian until the variational energy begins to stabilize and verify that the local observables exhibit the targeted VBC pattern. 
We then set the additional imprinting terms $\delta J$ and $h^{\prime}$ to zero and resume training under the original Heisenberg Hamiltonian, using  the same protocol as for the unbiased optimization runs.

\subsection{Optimization protocol at magnetization plateaus}\label{sec:App-protocol}

\begin{figure}[t]
    \centering
    \includegraphics[width=\columnwidth]{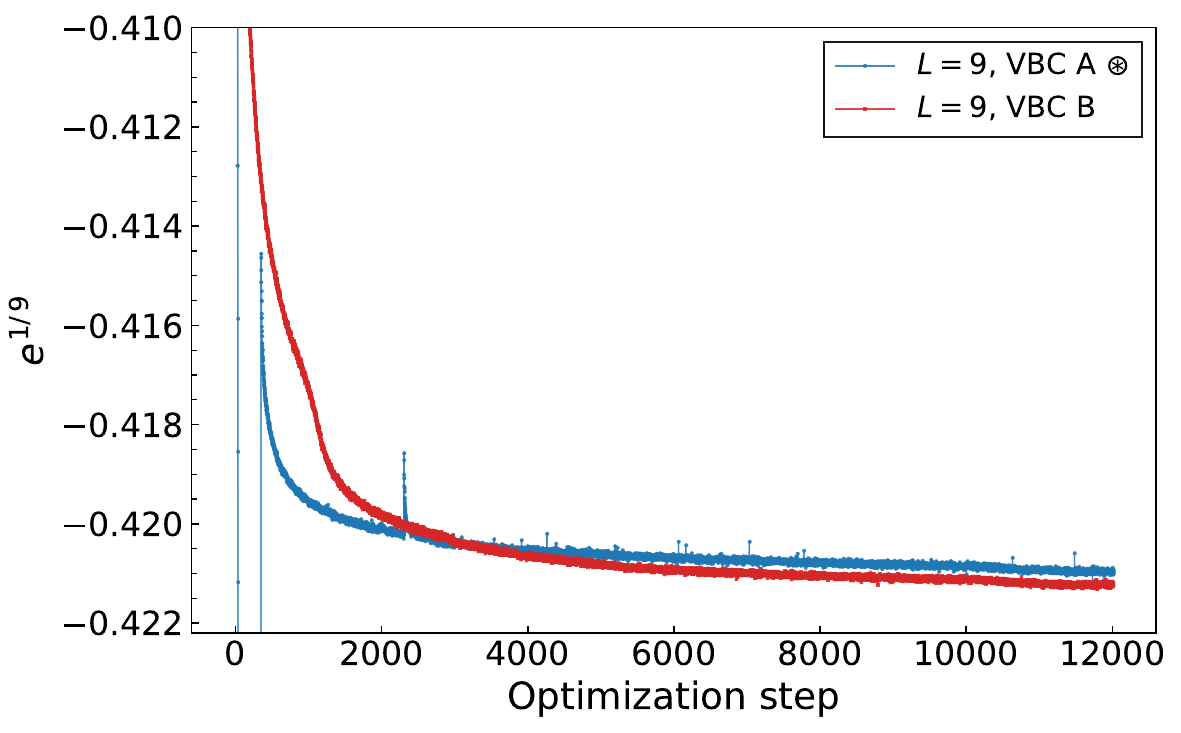}
    \caption{Variational energy per site throughout the optimization run for the two VBC states at the $m=1/9$ plateau for $L=9$. The VBC A state is obtained with imprinting (indicated with an asterisk), while the VBC B state is obtained with unguided optimization. The  discontinuous energy jump at the beginning of training  marks the transition from the perturbed imprinting Hamiltonian (lower energy, see Appendix~\ref{sec:App-imprinting}) to the original Heisenberg Hamiltonian, after which training resumes with free optimization. }
    \label{fig:optimization_curve}
\end{figure}

To obtain the variational ground states on the $m=1/9$, $1/3$, and $5/9$ magnetization plateaus, we keep the same ViT architectural hyperparameters  $(d, n_h,l)$ and SPRING momentum $\mu$ as described in Sec.~\ref{sec:implementation}, and we increase the number of Monte Carlo samples per step to $M=16384$.

For the $m=1/3$ and $m=5/9$ plateaus, we optimize for $5000$ steps using cosine-decaying schedules for the learning rate and the diagonal shift, with endpoints $(\tau_i,\tau_f)=(0.03,\, 5\times 10^{-3})$ and $(\lambda_i,\lambda_f)=(10^{-2},\, 10^{-4})$.

For the unguided runs on the $m=1/9$ plateau, we train for $12000$ steps with a piecewise schedule. Over the first $5000$ steps, we cosine-decay $\tau$ from $0.03$ to $0.015$ and $\lambda$ from $10^{-2}$ to $5\times 10^{-3}$, followed by an additional $5000$ steps keeping both fixed at their respective values. We then cosine decay $\tau$ down to $10^{-3}$ and $\lambda$ down to $10^{-4}$ over a final $2000$ steps.  
For runs initialized with imprinting, the fixed stage of the schedule is started earlier, since the initial imprinted VBC states are already at comparatively low variational energies. 

Representative optimization curves for both the imprinted and unguided protocols are shown in Fig.~\ref{fig:optimization_curve}, corresponding to the VBC A and VBC B states for $L=9$. We consistently find that the distinctive patterns of strong bonds that characterize the two VBCs can be observed in the first few thousand steps of training and remain unchanged thereafter. We thus expect that while longer runs or a different optimization protocol could conceivably lead to slightly improved final variational energies, they would likely not qualitatively change the nature of the observed states.

\bibliography{references.bib}

\end{document}